\documentclass{jfm}
\usepackage{float}
\floatstyle{plain}
\restylefloat{table}

\usepackage{graphicx}
\usepackage{verbatim}

\usepackage{subcaption} %
\usepackage{tgtermes}
\usepackage[T1]{fontenc}
\usepackage[svgnames]{xcolor}

\usepackage{amsmath}
\usepackage{amssymb}
\usepackage{newtxtext}
\usepackage{newtxmath}
\usepackage{natbib}
\usepackage{soul}

\usepackage{booktabs}
\usepackage{makecell}

\usepackage{hyperref}
\GetTitleStringSetup{expand}
\hypersetup{
    colorlinks = true,
    linkcolor  = blue,
    urlcolor   = blue,
    citecolor  = blue,
}

\newcommand{\ve}[1]{\ensuremath{\mbox{\boldmath$#1$}}}
\newcommand{\ma}[1]{\ensuremath{\mathbb{#1}}}

\usepackage{caption}
\usepackage{cancel}
\title{Inertial rotation of a small oblate spheroid in a simple shear flow}

\author{Z. Wang\aff{1}\footnote[1]{Email address for correspondence: z.wang5@tue.nl}, X. M. de Wit\aff{1}, D. Di Giusto\aff{2,3}, L. Bergougnoux\aff{3}, \'E. Guazzelli\aff{4},
C.\,Marchioli\aff{2}, B. Mehlig \aff{5}
\and F. Toschi\aff{1,6} }
\affiliation{
\aff{1}Department of Applied Physics and Science Education, Eindhoven University of Technology, 5600 MB Eindhoven, Netherlands
\aff{2} Dipartimento Politecnico di Ingegneria e Architettura, University of Udine, 33100 Udine, Italy
\aff{3} Aix-Marseille Universit\'e, CNRS, IUSTI, 13453 Marseille, France
\aff{4} Université Paris Cit\'e, CNRS, Mati\`ere et Syst\`emes Complexes, UMR 7057, 75013 Paris, France
\aff{5}Department of Physics, University of Gothenburg, 41296 Gothenburg, Sweden
\aff{6}CNR-IAC, Via dei Taurini 19, 00185 Rome, Italy
}

\begin{document}
\maketitle

\begin{abstract}
We compare experiments and fully-resolved particle simulations designed to match the experimental conditions of a weakly inertial, neutrally buoyant, moderately oblate spheroid in shear flow under confinement.
The study revisits an apparent discrepancy observed in earlier experiments, where a trajectory starting close to tumbling appeared to remain near the tumbling orbit, whereas weak-inertia theory predicts attraction to log-rolling for this particle shape.
Experimental and numerical results are benchmarked against theory valid for asymptotically small particle Reynolds numbers and for unconfined systems.
By accounting for confinement and inertia, together with sensitivity to initial conditions, and the finite observation time, we
 provide a consistent interpretation of the theoretical, experimental, and numerical results.
Specifically, we show that log rolling is the only stable long-time orbit for the moderately oblate spheroid considered here; neither confinement nor finite inertia changes this orbit selection. Instead, inertia tends to accelerate the drift toward log rolling, whereas confinement reduces it, with the confinement effect being dominant in the present parameter range.
\end{abstract}

\section{Introduction}
\label{sec:intro}

The rotation of non-spherical particles in shear flows is a central problem in fluid mechanics, with profound implications for both natural and industrial processes \citep{voth2017anisotropic,butler2018,MRV2026}.
The angular motion of such particles govern the bulk rheology of complex fluids and are critical in phenomena such as the settling of ice crystals in clouds, the dispersion of fibers in turbulence, and the behaviour of particulate suspensions in industrial applications.
Understanding these dynamics is essential for predicting the macroscopic properties of particle-laden flows and for optimizing processes in fields ranging from atmospheric science to materials engineering.

The foundational work on this problem was conducted by \citet{jeffery1922}, who analysed the angular dynamics of a spheroid in simple shear flow under the Stokes approximation and in the absence of Brownian motion.
\citet{jeffery1922} demonstrated that the axis of revolution of a spheroid follows one of an infinite one-parameter family of closed periodic orbits, now known as Jeffery orbits.
These orbits are characterized by their dependence on initial conditions, rendering the inertialess limit indeterminate.
\citet{bretherton1962} later extended these results to bodies of revolution by introducing an effective aspect ratio, thereby generalizing the findings of \citet{jeffery1922} to a broader class of particle shapes.

A key issue raised by \citet{jeffery1922}'s analysis is the indeterminacy of the particle final state in the inertialess limit.
For an isolated particle, in the absence of Brownian motion, interactions with other particles, or other perturbations, the specific orbit adopted by the particle depends solely on its initial orientation.
This indeterminacy has two significant consequences. First, it  leaves the long-time orbit undetermined by the particle shape and flow alone, because the orbit constant remains set by the initial condition. Second, it suggests that even small perturbations, such as orientational diffusion or minor shape asymmetries, can dramatically alter particle dynamics.
For very small colloidal particles, orientational diffusion regularizes the problem by breaking the degeneracy of the Jeffery orbits, enabling the calculation of effective viscosities \citep{hinch1972}.
For larger particles, inertial effects eliminate the degeneracy. The question is whether this results in a unique distribution of particle orientations and steady-state rheology in simple shear flow, thereby eliminating rheological degeneracy.

\citet{jeffery1922} hypothesized that weak inertial effects would drive particles toward orbits of minimum dissipation, corresponding to log-rolling (also referred to as spinning) for prolate spheroids and tumbling for oblate spheroids.
Early experimental investigations by \citet{taylor1923} and \citet{trevelyan1951} were inconclusive.
The first analytical study of weak fluid inertia effects on nearly spherical particles in simple shear flow was conducted by \citet{saffman1956}, which appeared to support \citet{jeffery1922}'s hypothesis.
However, subsequent experiments by \citet{karnis1966} with rods and disks observed particles migrating toward orbits of maximum energy dissipation, contradicting \citet{saffman1956}'s findings.

The role of weak inertial effects has since been addressed in theoretical works, including those by \cite{subramanian2005} for slender rods and by \citet{einarsson2015b, einarsson2015a} for spheroids of arbitrary aspect ratio, derived later also by \citet{dabade2016effect,marath2017, marath2018}.
These studies conclude that fluid and particle inertia break the degeneracy of Jeffery orbits.
Specifically, slender-body theory predicts that tumbling is stable for slender fibres \citep{subramanian2005}, a finding that has also been evidenced and extended to prolate spheroids in numerical simulations \citep{qi2003,you2007,huang2012,rosen2014,mao2014motion}.
\citet{einarsson2015b, einarsson2015a} computed the stability of Jeffery orbits under inertial perturbations for spheroids of arbitrary aspect ratio to the order of the particle Reynolds numbers using the reciprocal theorem and symmetry arguments, coming to the opposite conclusion compared with Saffman.
Their findings indicate that tumbling is stable and log-rolling is unstable for prolate spheroids, while the opposite is true for oblate spheroids, provided they are not too thin.
Additionally, a new bifurcation was identified, whereby an unstable limit cycle appears for very thin oblate spheroids with an aspect ratio smaller than 0.14 \citep{einarsson2015b}, leading to the stability of both tumbling and log-rolling orbits.
\citet{dabade2016effect} came to the same conclusion using a slightly different method based on a spheroidal harmonics formalism.
These findings were confirmed by direct numerical simulations \citep{rosen2015numerical}, which further explored the impact of flow confinement on the stability of tumbling and spinning orbits.
In summary, inertial corrections drive prolate spheroids toward the tumbling orbit, whereas the situation is more complex for oblate spheroids.
Oblate spheroids are attracted to either the sole log-rolling orbit or both the vorticity-aligned and tumbling orbits, depending on whether their aspect ratio is larger or smaller than a critical value of approximately 0.14.

Recent experiments by \citet{digiusto2024influence} have provided quantitative insights into the influence of inertia on the stability of Jeffery orbits.
These experiments confirmed an irreversible drift across Jeffery orbits toward attracting limit cycles, in qualitative agreement with the theory discussed above.
For prolate particles, a unique tumbling motion in the flow-gradient plane was observed, regardless of initial orientation and aspect ratio.
Oblate particles, however, exhibited a more complex behaviour, drifting toward either the log-rolling or tumbling orbit depending on their initial orientation.
The only notable difference between the experiments of \citet{digiusto2024influence} and the theoretical prediction, as well as the simulations of \citet{rosen2015numerical}, is the apparent stability of tumbling orbits for all oblate particles.
The stability of tumbling-like trajectories has also been found for other oblate particles, such as rings, by \citet{digiusto2025}.
Although theories predict a bifurcation towards a single stable log-rolling orbit above a critical aspect ratio of approximately 0.14,
one experimental trajectory (RUN10 in figure 11 of \cite{digiusto2024influence}) for the oblate spheroid with aspect ratio $\lambda\simeq0.56$ appeared to remain close to the tumbling orbit.
This discrepancy may be attributed to the need for longer observation periods to detect the drift towards log-rolling in the experiments, as noted by \citet{digiusto2025}. Also, the theory is valid for asymptotically small particle Reynolds numbers and for unconfined systems which are conditions not easily met in experiments.

There remains a need for a more thorough explanation of the discrepancies between experiments and theoretical predictions. This is precisely the objective of the present paper, where we propose a combination of new experimental observations and direct numerical simulations based upon the immersed boundary method (IBM). The results are then compared with existing theoretical models in order to resolve the aforementioned discrepancies, focusing on the effects of inertia and confinement on the angular dynamics and stability of oblate spheroids in shear flows. The experimental and numerical methods are detailed in \S\,\ref{sec:methods}, while \S\,\ref{sec:results} presents a comparison of experimental, numerical, and theoretical results. Conclusions are drawn in \S\,\ref{sec:conclusion}.

\begin{figure}   \centerline{\includegraphics[width=0.99\linewidth]{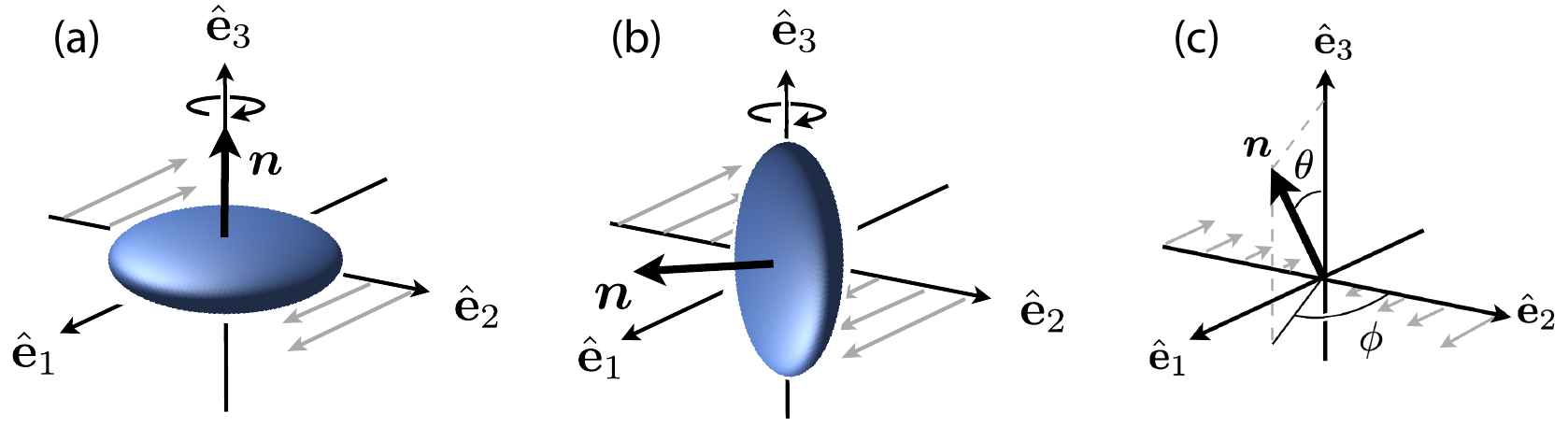}}
    \caption{\label{fig:schematic} Oblate spheroid in a simple shear.
    ({a}) log-rolling (LR) orbit, ({b}) tumbling (T) in the flow-gradient plane.
    The particle-symmetry axis is denoted
    by $\ve n$, and the flow vorticity by $\ve \Omega$.
    The basis unit vectors of the lab coordinate system are denoted by $\hat {\bf e}_1$, $\hat {\bf e}_2$, and $\hat {\bf e}_3$,
    where $\hat{\bf e}_1$ is the flow direction, $\hat{\bf e}_2$ is the velocity-gradient direction, and $\hat{\bf e}_3$ is the vorticity direction. The flow-gradient plane is the $(\hat{\bf e}_1,\hat{\bf e}_2)$ plane and the flow-vorticity plane is the $(\hat{\bf e}_1,\hat{\bf e}_3)$ plane. The spheroid semi-axis perpendicular to $\ve n$ is $a$, and the semi-axis parallel to $\ve n$ is $b$.
    ({c}) Definitions of the azimuthal and polar angles, $\phi$ and $\theta$.
   }
\end{figure}

\section{Methods}
\label{sec:methods}

\subsection{Problem description}

We consider a neutrally buoyant oblate spheroid at the centre of a simple shear flow as depicted in figure\,\ref{fig:schematic} showing (a) log-rolling (LR) where the particle-symmetry vector, $\ve n$, is aligned with vorticity and (b) tumbling (T) where $\ve n$ lies in the flow-shear plane.
The components of the flow velocity $\mathbf{u}$ in the laboratory coordinate system are $u_1 = \dot{\gamma} x_2, u_2=u_3=0$, where $\dot{\gamma}>0$ is the shear rate.
The matrix $\ma A$ of fluid-velocity gradients is thus $\ma A = [[0,\dot{\gamma},0],[0,0,0],[0,0,0]]$.
Its symmetric and antisymmetric parts are denoted by $\ma S$ and $\ma O$.
The elements of $\ma O$ are related to the components of the vorticity vector $\boldsymbol{\Omega} = \nabla \times \mathbf{u}$ as $O_{ij} = -\frac{1}{2}\epsilon_{ijk}\Omega_k$, where $\epsilon_{ijk}$ is the completely asymmetric rank-3 tensor.
For the given simple shear flow, $\mathbf{u} = (\dot{\gamma} x_2, 0, 0)$, the only non-zero component of vorticity is $\Omega_3 = -\dot{\gamma}$.
In short, $\ve \Omega$ aligns with the $\hat{\mathbf{e}}_3$-axis.
We define the particle aspect ratio $\lambda$ as the ratio of its half-length along the symmetry axis, $b$, to its half-length perpendicular to that axis, $a$. Thus, $\lambda>1$ corresponds to prolate particles, whereas $\lambda<1$ corresponds to oblate particles.
Experiments and simulations are performed in a slab of finite width $L$, and this confinement is characterised by the non-dimensional parameter, $\kappa = 2a/L$.
The fluid and particle inertia are characterized by the particle Reynolds number Re$_p = \rho_f a^2 \dot{\gamma} /\mu$ and the Stokes number St$=\rho_p a^2 \dot{\gamma} /\mu = \text{Re}_p \rho_p/\rho_f$, where $\rho_p$ and $\rho_f$ are the particle and fluid densities, respectively, and $\mu$ is the fluid dynamic viscosity. In this study, the particle is neutrally buoyant (i.e., $\rho_p=\rho_f$), so that $\text{Re}_p=\text{St}$.

\subsection{Experiments}

Experiments are performed in a linear shear cell (500 mm long, 40 mm wide, and 90 mm deep) at a constant shear rate $\dot{\gamma}=3.1~\mathrm{s}^{-1}$.
The cell is filled with a solution of pure water and citric acid to achieve isodensity conditions, as well as Ucon oil to adjust the final viscosity and control the effect of fluid inertia.
Further details on the preparation and characterisation of the working fluid, including the properties of the constituents and the density and viscosity measurements, are provided in Appendix\,\ref{sec:working_fluid}.
We consider an oblate spheroidal particle with an aspect ratio of $0.56 \pm 0.01$, which has been manufactured using 3D-printing.
For each experimental run, the particle is suspended in the middle of the confined volume, and sheared through a flexible, transparent plastic belt.
Further details on the preparation of the initial particle orientation and the measurement uncertainty relevant to the initial condition are provided in Appendix\,\ref{sec:initial_orientation}.
Two cameras with perpendicular axes record the particle images, capturing both the $\hat {\bf e}_1-\hat {\bf e}_2$ (flow-gradient) and $\hat {\bf e}_1-\hat {\bf e}_3$ (flow-vorticity) planes.
Experiments terminate when the spheroid leaves the camera fields of view,
and therefore the present analysis uses the time interval over which the particle remains in the field of view and does not measure cross-stream migration.
Further details regarding the experimental protocol can be found in \citet{digiusto2024influence}.
In addition to the experiments reported in \citet{digiusto2024influence}, the present study considers much longer runs, up to 1.6 times.
This is significant because the decay rate toward LR is small at small Re$_p$ and therefore the particle rotation experiences a slow initial evolution. Consequently, the final stable state must be determined through the analysis of extensive experimental time series.

The particle orientations ($n_1, n_2, n_3$) are estimated from the experimental video recordings as presented in \citet{digiusto2024influence}.
First, gradient-thresholding methods are applied to measure the particle projected lengths in both camera planes and to reconstruct the three-dimensional Axes-Aligned Bounding Boxes (AABBs).
These are assembled combining the projected $\hat {\bf e}_1$ and $\hat {\bf e}_2$ lengths observed by the flow-gradient camera with the $\hat {\bf e}_3$ extension observed in the flow-vorticity plane, opportunely rescaled to exactly match the resolution on the suspended particle.
Finally, the AABBs are normalized by dividing by the particle diameter, measured in pixels.
Projected lengths are then stacked and processed by a dense neural network to infer the particle orientation vector, one single pair of frames at a time.
The neural network is made of three layers with 32 neurons each, and a final output layer returning the three components of the particle orientation vector, determining 2,339 trainable parameters.
The Rectified linear unit is set as the activation function for all the layers except the last one, where linear mapping is adopted.
Differently from \citet{digiusto2024influence}, we select the cosine distance between the true and predicted particle orientation vectors as the loss function to be minimized in the model training: $\mathscr{L}=1.0-{\ve n}_{pred}\cdot {\ve n}_{true}$.
A supplementary penalty on negative $n_3$ values is also added to the loss function.
The neural network is trained with a learning rate $10^{-3}$ during 500 epochs in a supervised framework on 20,000 synthetic particle orientations.
Training data are generated by exploiting the known geometrical relations that facilitate the calculation of the particle projected lengths from a large set of imposed particle orientations, as described in \citet{digiusto2024influence}.
The robustness of this approach is enhanced in this study by introducing zero-mean random Gaussian noise in the training data.

A statistical approach is adopted to characterise the uncertainties in the measurements.
By repeatedly perturbing 1,000 randomly chosen experimental recording frames with Gaussian noise, a sub-pixel standard deviation uncertainty on the projected length measurements in both camera fields is estimated.
Consequently, random outliers in the reconstructed orbits can be most plausibly attributed to spurious visual artefacts, such as bubbles, dirt in the fluid, or imperfections in the transparent plastic belt.
Uncertainties associated with the neural network are also evaluated.
The assessment of epistemic uncertainty is conducted through the analysis of 1,000 randomly selected AABB measurements.
This is achieved by implementing repeated inferences of the particle orientation vector with activated neurons dropout, a process which is found to be negligible.
Aleatoric uncertainty is then evaluated by modelling the previously estimated uncertainty distributions of the AABBs measurements as normal distributions and repeatedly propagating them through the Neural Network.
Regardless of the orientation of the particle, the uncertainty in the inference of $n_3$ is found to correspond to a standard deviation $\sigma_{n_3} = 0.06$.
It should be noted that this value is contingent on the number of trainable parameters that constitute the Neural Network.

\subsection{Simulations}

The angular dynamics of the spheroid in a shear flow is numerically investigated by the fully resolved IBM.
The fluid domain is represented by a fixed, Cartesian grid on which the incompressible Navier–Stokes equations are solved.
Immersed particles are represented by Lagrangian markers on a separate mesh, with the no-slip boundary condition enforced via a distributed forcing term added to the momentum equations. Fluid and particle dynamics are fully coupled through the Newton–Euler equations governing translational and rotational motion. Time integration is performed using a three-step Runge–Kutta scheme for advection and particle dynamics, while fluid diffusion is treated using the Crank–Nicolson method. Incompressibility is maintained by solving a pressure Poisson equation at each time step.
Pressure and momentum are coupled via a fractional-step method. This IBM framework has been extensively described and validated \citep{rai1991direct, verzicco1996finite, de2016moving, piumini2024particle, witkamp2024low,huseby2026bifurcations}.
Here, we consider a neutrally buoyant spheroid in a shear flow generated by imposing equal and opposite velocities on the top and bottom boundaries. The gradient direction of the flow is confined in a slab of width $L$ with no-slip boundary conditions.
In the flow- ($\hat{\bf{e}}_1$) and vorticity- ($\hat{\bf{e}}_3$) directions, periodic boundary conditions are applied.
    The corresponding periodic lengths are chosen independently of the wall-normal gap and are characterised by a lateral domain factor $f$, defined such that $f=1$ corresponds to the cubic domain. The confinement parameter $\kappa$ therefore characterises only the wall-normal confinement, consistently with its definition in the experiments, whereas $f$ controls the distance to the periodic images in the flow and vorticity directions.
    A systematic convergence study with respect to $f$ is reported in Appendix\,\ref{sec:f}. This test keeps the wall-normal distance and the particle size fixed, while increasing only the periodic lengths. It therefore isolates the influence of periodic images in the directions with periodic boundary conditions from the finite-domain effect measured by the wall-normal confinement parameter $\kappa$.
    As shown in Appendix\,\ref{sec:f}, increasing the lateral periodic lengths slightly changes the quantitative decay rate $\gamma_{\rm LR}$ and does not alter the selected long-time orbit.
    Based on this test, we use $f=2$ for the simulations directly compared with the experiments, and $f=4$ for the quantitative parameter scans in Figs.\,\ref{fig:drift} and \ref{fig:DCmap}, in order to minimise the influence of periodic images while retaining feasible computational cost.
    In the following, ``confinement'' refers specifically to the wall-normal confinement characterised by $\kappa$, consistent with the experimental definition, whereas the lateral periodic extent is reported independently through $f$.
    The detailed parameter settings in the simulations are reported in Appendix \ref{sec:sim_param}. The production runs use a uniform isotropic grid with
    $D_{eq}/\Delta x \simeq 42.2$ ($\approx 51$ grid cells across the particle major diameter $2a$). A grid-refinement check for the experiment-matched case is reported in Appendix~\ref{sec:grid_refinement_check}.
    The centre of the spheroid remains fixed at the mid-plane to isolate the angular dynamics from possible translational inertial migration of the particle centre position.
Previous theoretical and numerical results establish that the mid-plane is a stable equilibrium position for neutrally buoyant spheres at sufficiently low channel Reynolds numbers. Whether this conclusion applies quantitatively to a three-dimensional oblate spheroid has not been established, and translational migration is not examined here (see Appendix\, \ref{sec:inertial_mig}).

\subsection{Theory}

The theory of \citet{einarsson2015a} relies upon symmetry considerations to derive the form of the angular equation of motion for a neutrally buoyant spheroid in an unbounded simple shear to linear order in Re$_p$,
 \begin{align}
  \dot n_i =\epsilon_{ijk} \omega_j n_k &+  \beta_1 (n_pS_{pq}n_q)\left(\delta_{ij}-n_in_j\right)S_{jk}n_k  + \beta_2 (n_pS_{pq}n_q)O_{ij}n_j  \nonumber\\
            &+ \beta_3 \left(\delta_{ij}-n_in_j\right) O_{jk}S_{kl}n_l + \beta_4 \left(\delta_{ij}-n_in_j\right) S_{jk}S_{kl}n_l\,,
            \label{eq:einarsson}
\end{align}
where repeated indices are summed over.
The first term on the r.h.s. is Jeffery's equation of motion with Jeffery's angular velocity $\ve \omega$ with components $\omega_j = \Omega_j + \Lambda \epsilon_{jkm}n_k n_l S_{ml}$.
The remaining terms are inertial corrections.
The equation is non-dimensional, where lengths were made dimensionless using the particle size $a$, and time with the shear rate $\dot{\gamma}$.
The parameter $\Lambda$ is defined as $\Lambda = (\lambda^2-1)/(\lambda^2+1)$.
Note that equation\,\eqref{eq:einarsson}
 assumes only symmetry and that corrections are of second order in $\ma S$ and $\ma O$.  The coefficients $\beta_\alpha$ depend on the shape of the particle.
Using the reciprocal theorem, \citet{einarsson2015a} determined the numerical values of the non-dimensional coefficients for spheroidal particles.
In the dimensional form of equation\,\eqref{eq:einarsson}, $\dot n_i, \omega_j, S_{pq}$, and $O_{ij}$ have units of inverse time, whereas the coefficients $\beta_\alpha$ have units of time.
Note that, in the comparison with the theory of \citet{einarsson2015a} made by \citet{digiusto2024influence} in their appendix B.1, dimensionless coefficients $\beta_\alpha$ were mistakenly used in the dimensional form of equation\,\eqref{eq:einarsson}, which partially affected the plotted theoretical curves without changing the main conclusions, see \cite{di_giusto_corrigendum_2025} \citep[Corrigendum of][]{digiusto2024influence} for details.

\citet{dabade2016effect} derived the same results as \citet{einarsson2015a}, using different basis functions to represent the spheroidal particle. In addition, they quoted an explicit expression for the normalised change of the orbit constant due to inertia [their Eq. (5.19)], see also Appendix B.2 of Di Giusto et al. (2024). This change is examined in \S\,\ref{subsec:influence} to characterize the relative importance of confinement and inertia on the drift of the spheroid from the Jeffery orbits.

\section{Results}
\label{sec:results}

In this section, we present the key findings from our comparative study.
We first examine the angular motion of the spheroid by looking at the time evolution of the components ($n_1$, $n_2$, $n_3$) of its orientation vector $\ve n$.
Then, we discuss the dependence of the resulting long-term rotational behaviour on the particle Reynolds number and on the stability exponent of the LR orbit, $\gamma_{\text{LR}}$, which characterises the exponential decay rate of the orbit constant $C$ through the relation $C \sim exp(-\gamma_{\text{LR}} t \dot{\gamma})$.
In particular, we quantify this dependence by deriving a scaling law for the orbit constant $C$.
Finally, we discuss the effects of confinement and inertia by looking at the change in the orbit constant in a single Jeffery period, which can be conveniently introduced to characterize the orbit drift.
We remark here that both experiments and simulations have been conducted considering a wall-bounded domain in which confinement has an effect on particle rotation.
Such an effect is not present in the theoretical predictions, which, however, share exactly the same initial conditions as the simulations.

\begin{figure} \centerline{\includegraphics[width=0.99\linewidth]{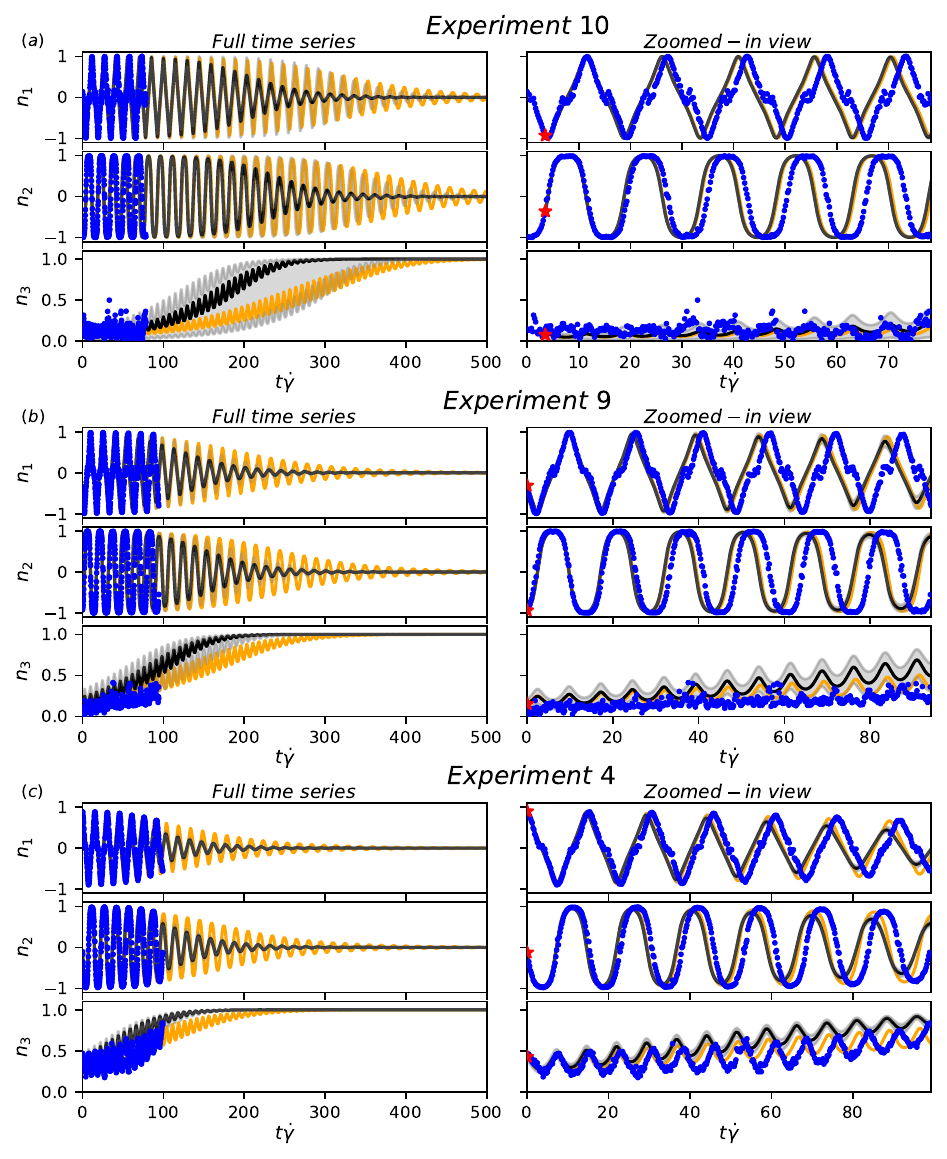}}
    \caption{\label{fig:comparison} Comparison of the experimental time series (blue symbols) with the results of numerical simulations (orange solid lines) and theoretical predictions of Eq.~(\ref{eq:einarsson}) (black solid lines) for the case of spheroids with Re$_p=$ 0.43, $\lambda=0.56$ and $\kappa=0.2$.
    The only difference among the different runs is the initial orientations of the particle.
    The plots in panel ({a}) refer to an experimental run (labeled here as {\tt{Experiment 10}} to follow the notation used in the Jupyter notebook).
    The plots in panel ({b}) refer to the experimental run labeled {\tt{Experiment 9}} in the Jupyter notebook.
    The plots in panel ({c}) refer to the experimental run labeled {\tt{Experiment 4}} in the Jupyter notebook.
    Left-hand panels show the full time evolution of particle orientation, while right-hand panels show a zoomed-in view to highlight the different behaviour of the curves.
    The grey shaded area indicates the theoretical predictions obtained with initial conditions perturbed by $n_3 \pm \sigma_{n_3}$, and is shown here to highlight the sensitivity of the theory to the initial orientation of the particle.
    The red stars in the zoomed-in panels mark the experimental orientation used to initialize the corresponding theory and simulation curves.
    See supplementary materials for the directory of the figure including the data and the Jupyter notebook, \href{https://cocalc.ai/share/notebook/figure_2/Figure_2.ipynb}{https://cocalc.ai/share/notebook/figure\_2/Figure\_2.ipynb}.
    Panels (a), (b) and (c) correspond to experimental runs 10, 9 and 4, respectively, as labelled in the Jupyter notebook, and to Movie 1, Movie 2 and Movie 3, respectively.
    }
\end{figure}

\subsection{Time evolution}
\label{subsec:timeevolution}

Figure \ref{fig:comparison} shows a comparison between experimental time series (blue symbols), ab-initio simulations (orange solid lines), and theoretical predictions obtained by numerically integrating equation\,\eqref{eq:einarsson} (black solid lines).
The time series refer to different experimental runs conducted for the case of a spheroid with Re$_p$ = 0.43, $\lambda=0.56$ and $\kappa=0.2$, which is chosen here as reference for the discussion.
The difference between each run is just the initial orientation of the particle, namely the initial values of ($n_1$, $n_2$, $n_3$).
Additional time series for the same particle geometry and size are provided in Appendix\,\ref{sec:app_data} as well as in the Supplementary Material, where the directory of the figure including the data and the Jupyter notebook is given.
The grey shaded area around the black lines indicates the theoretical predictions obtained when the initial conditions are perturbed by $n_3 \pm \sigma_{n_3}$, and is shown to highlight the sensitivity of the dynamics to small changes in the initial orientation imposed on the spheroid.
The two boundaries of the shaded region (gray solid lines) correspond to the fastest and slowest convergence cases.
The left-hand panels of figure\,\ref{fig:comparison} refer to the entire time evolution of $n_1$, $n_2$, and $n_3$, which covers $t\dot{\gamma}=500$ dimensionless time units, for the same particle geometry but three different initial orientations.
The right-hand panels provide a zoomed-in view of the first 100 time units, to highlight the different behaviour of the curves.

Overall, experiments, simulations, and theory are able to qualitatively capture the oscillating and decaying behaviour of $n_1$ and $n_2$.
As far as the time evolution of $n_3$ is concerned, the theoretical predictions and numerical simulations show a consistent increase of $n_3$ towards unity from very small values and thus exhibit a general tendency to drift towards the LR orbit. This indicates that LR is stable while T is unstable.
To quantify the difference in rotation frequency, we extracted the rotation period $T_r$ from the experimental and numerical time series and compared it with the Jeffery period $T_J$, which is defined as $T_J=2\pi(\lambda+1/\lambda)/\dot{\gamma}$. The results are reported in Appendix\,\ref{sec:Tj_Tr}.
The oscillation frequency predicted by the theory is always slightly higher than that measured in experiments, and simulation results generally fall in between.
The same analysis shows that the trajectories remain approximately Jeffery-like over each rotation: the root-mean-square angular deviations from the corresponding best-fitting Jeffery orbits are below $1^\circ$ for the simulation and theory and approximately $2.4^\circ$ for the experiments (Appendix \,\ref{sec:Tj_Tr}).
A more substantial difference can be seen from the convergence rate, which appears to be always faster for the theory compared to the simulations. This can be attributed to the presence of confinement effects in the latter, as we shall show hereafter. Likewise, confinement can also explain the observed reduction in oscillation frequency when comparing the simulation and the theory, which is found to decrease by approximately 8\% as the confinement $\kappa$ increases from 0.2 to 0.7 in the numerical simulations.

The drift of the orbit is not always observed within the time window available for the experiments (see figure\,\ref{fig:comparison}({a}) and figure\,\ref{fig:comparison}({b}), for instance), which shows a much slower drift towards LR when the initial condition is close to the T orbit.
In this case, the experimental measurements appear to be too short to definitively confirm that T is unstable.
The evolution of $n_3$ also highlights a strong dependence on the initial orientation, which is not observed for $n_1$ and $n_2$.
For example, in figure\,\ref{fig:comparison}({b}), the convergence rate is found to change by as much as 30\% when the initial value of $n_3$ is changed by an amount equal to $\sigma_{n_3}$.
This shows that even small perturbations in the initial orientation can significantly affect the predicted relaxation dynamics, leading to experimental trajectories that may lie outside of the grey shaded region associated with the theoretical predictions.
It should be noted that the experimental measurements of the three components exhibit increased noise at values close to zero.
These fluctuations reflect unavoidable uncertainties associated with the image-based orientation reconstruction.
In contrast, macroscopic disturbances such as bubbles, dirt, or imperfections of the plastic belt primarily contribute to the more distant scattered points observed in the data.

\subsection{Influence of confinement and inertia}
\label{subsec:influence}

\begin{figure}
\centerline{\includegraphics[width=1\linewidth]{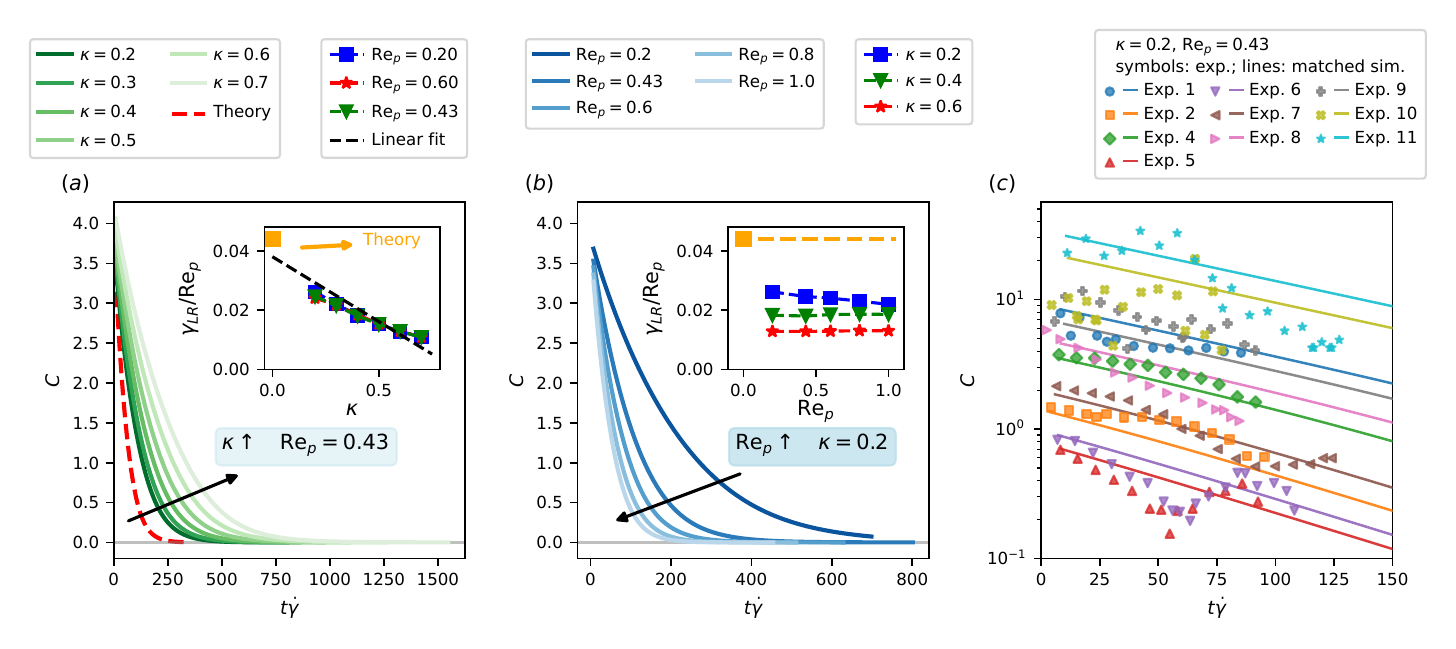}}
    \caption{\label{fig:drift} Quantification of the orbit drift. Panel
    ({a}): drift of the orbit constant $C$ as a function of time for different values of $\kappa$ for Re$_p=0.43$ (simulation results with the experimental setting).
    The dashed red line shows the theoretical prediction obtained by numerical integration of Eq.~(\ref{eq:einarsson}) for the same particle geometry and with the same initial particle orientation as in {\tt Experiment 4} (see figure\,\ref{fig:comparison}).
    The orange square in the inset is the corresponding value of $\gamma_{\rm LR}/\text{Re}_p$.
    A systematic scanning of the dependence of $\gamma_{\rm LR}/\text{Re}_p$ on $\kappa$ for Re$_p=0.30$ (squares), Re$_p=0.43$ (triangles, experimental setting), and Re$_p=0.60$ (stars), is shown in the inset. The black dashed line represents a linear fit through the simulation data and the unbounded-theory value at $\kappa=0$, $\gamma_{\text{LR}}/\text{Re}_p = -0.0437 \kappa + 0.0380$.
    Panel ({b}): drift of the orbit constant $C$ as a function of time for different values of Re$_p$ for $\kappa =0.2$ (simulation results with the experimental setting).
    A systematic scanning of the dependence of $\gamma_{\rm LR}/\text{Re}_p$ on Re$_p$ for $\kappa=0.2$ (squares, experimental setting), $\kappa=0.4$ (triangles), and $\kappa=0.6$ (stars) is shown in the inset, where
    the orange dashed line represents the unbounded weak-inertia value $\gamma_{\text{LR}}/\text{Re}_p = 0.0439$,  obtained by applying the same fitting procedure to the theoretical $C(t)$.
    Within the weak-inertia theory, $\gamma_{\text{LR}}$ is linear in
    Re$_p$ at leading order, so that the normalised decay rate
    $\gamma_{\text{LR}}/\text{Re}_p$ is independent of Re$_p$ and is therefore
    shown as a horizontal reference line over the displayed Re$_p$ range.
    Panel ({c}) compares each experimental run with its own matched simulation (same colour); all ten runs with different initial orbit constants $C$, decay toward log-rolling, and each experiment follows the decay of its matched simulation. The vertical axis is shown on a logarithmic scale.
    See supplementary materials for the directory of the figure including the data and the Jupyter notebook,
    \href{https://cocalc.ai/share/notebook/figure_3/Figure_3.ipynb}{https://cocalc.ai/share/notebook/figure\_3/Figure\_3.ipynb}.
    }
 \end{figure}

To verify the relative importance of confinement and inertia effects on the stability of the LR and T orbits, we next examine the time evolution of the orbit constant $C$ \citep{hinch1979rotation,leal1971effect}, defined such that the Jeffery orbits are of the form:
\begin{subequations}\label{Jefferey}
\begin{align}
\tan \phi &= \lambda \tan \left[ \frac{t\dot{\gamma}}{\lambda + (1/\lambda)}\right ] \, ,
\label{Jeffery1}\\
\tan \theta &= \frac{C \lambda}{ (\lambda^2\cos^2 \phi +  \sin^2 \phi)^{1/2}} \, ,
\label{Jeffery2}
\end{align}
\end{subequations}
with $n_1 = \sin \theta \sin \phi$, $n_2=\sin \theta \cos \phi$, and $n_3 = \cos \theta$ (the definitions of the azimuthal and polar angles, $\phi$ and $\theta$, are  indicated in Fig.\,\ref{fig:schematic}({c})).
Note that LR-orbits are characterized by $C=0$, while T-orbits are characterized by $C=\infty$.
Appendix\,\ref{sec:Ccalc} yields more details on the procedure adopted in the present study to calculate $C$.
To single out the effect of confinement, the time evolution of $C$ as predicted by theory (dashed line) is compared to the simulation results (shaded green lines) in panel (a) of figure\,\ref{fig:drift}, for different values of $\kappa$ when Re$_p=0.43$.
The inset in this panel shows the behaviour of the stability exponent as a function of $\kappa$ when rescaled by Re$_p$.
To single out the effect of inertia, in panel (b) of Figure\,\ref{fig:drift}, we show the time evolution of $C$ for different values of Re$_p$ when $\kappa=0.2$.
In all cases, the initial particle orientation is the same as in {\tt Experiment 4} (figure\,\ref{fig:comparison}(c)), for which the agreement between simulations and experiments is best.

Figure\,\ref{fig:drift} confirms that the orbit drift is influenced by both confinement, which acts to decrease $\gamma_\textrm{LR}$ at increasing $\kappa$ (see figure\,\ref{fig:drift}({a}) and its inset), and inertia, which acts to increase $\gamma_\textrm{LR}$ at increasing Re$_p$ (see figure\,\ref{fig:drift}({b}) and its inset).
The corresponding experimental estimates of orbit drift are shown in Fig.\,\ref{fig:drift}(c), where they are superimposed on the numerical curves (analysis methodology is reported in Appendix\,\ref{sec:Ccalc}). Although the experimental records cover a shorter time window and start from different initial values of $C$, they show the same overall trend: $C$ decreases with time, indicating drift away from tumbling and toward LR. This comparison supports the interpretation that the apparent tumbling-like behaviour in short experiments is caused by slow drift and finite observation windows rather than by a stable tumbling orbit.
In other words, confinement leads to a slower drift towards a stable LR-orbit whereas inertia favours a faster drift.
In a fully resolved IBM simulation, approaching the unbounded limit by directly reducing $\kappa$ is computationally expensive because the fluid domain must be enlarged while maintaining the particle resolution. Therefore, instead of attempting to simulate $\kappa=0$ directly, we vary $\kappa$ over an accessible range and assess the asymptotic trend as the unbounded limit is approached.
The inset of figure\,\ref{fig:drift}({a}) shows that the stability exponent of the LR-orbit $\gamma_{\rm LR}$, calculated as explained in Appendix\,\ref{sec:Ccalc}, exhibits a clear scaling behaviour with respect to Re$_p$, such that $\gamma_{\rm LR}/\text{Re}_p$ collapses on a single value for a given confinement $\kappa$. This trend is consistent with that obtained by \citet{rosen2015numerical} for much smaller particle Reynolds number, Re$_p= 2.5\times 10^{-4}$. Furthermore, the inset shows that, as $\kappa$ decreases, the simulation trend is directed toward the unbounded
theoretical value (orange square). A linear fit through the simulation data and the unbounded-theory value at $\kappa=0$ yields the relation $\gamma_{\text{LR}}/\text{Re}_p = -0.0437 \kappa + 0.0380$ (black dashed line).
This implies that all the curves shown in figure\,\ref{fig:drift}({a}) collapse onto a single curve when time is rescaled by Re$_p^{-1}$, namely when $C$ is plotted against $t \dot{\gamma} \times \text{Re}_p$ (see the Jupyter Notebook).
Note that the linear fit is valid for weakly inertial particles. Indeed, at the limiting value Re$_p = 0$, no orbit drift occurs and hence $\gamma_{LR} = 0$.
An important finding is that the linear relation between $\gamma_{\text{LR}} / \text{Re}_p$ and $\kappa$ incorporates both confinement and inertia effects on $\gamma_{\text{LR}}$ and can be used to make proper comparisons between theory, which was derived for weak-inertia particles in unbounded shear flow, and experiments, which are inevitably affected by flow confinement and are typically performed with weakly inertial particles. It is useful to distinguish the orbit-drift correction considered here from other weak-inertia particle problems in shear flow. In classical shear-induced lift problems, the leading inertial contribution may arise from a singular outer region of size $\ell_s=\sqrt{\nu/\dot{\gamma}}$, the Saffman length \cite{saffman1965lift,candelier2023second}. Similar outer-region effects are also known to determine the first non-zero correction to the rotation period of spheroids \cite{meibohm2016angular,marath2017effect}, which occurs at $O(\text{Re}^{3/2})$ for aspect ratios of order unity. If the present orbit drift were controlled by such a singular outer region, the confinement dependence would be expected to involve the wall distance measured in units of $\ell_s$, namely $(L/2)/\ell_s=\sqrt{Re_p}/\kappa$, rather than $\kappa$ alone. The collapse of $\gamma_{LR}/Re_p$ at fixed $\kappa$ for different $Re_p$ suggests that the confinement effect discussed here is more naturally interpreted as a finite-domain modification of the leading $O(Re_p)$ orbit drift, rather than as a Saffman-length-controlled outer-region correction.

\begin{figure}
    \centerline{\includegraphics[width=1\linewidth]{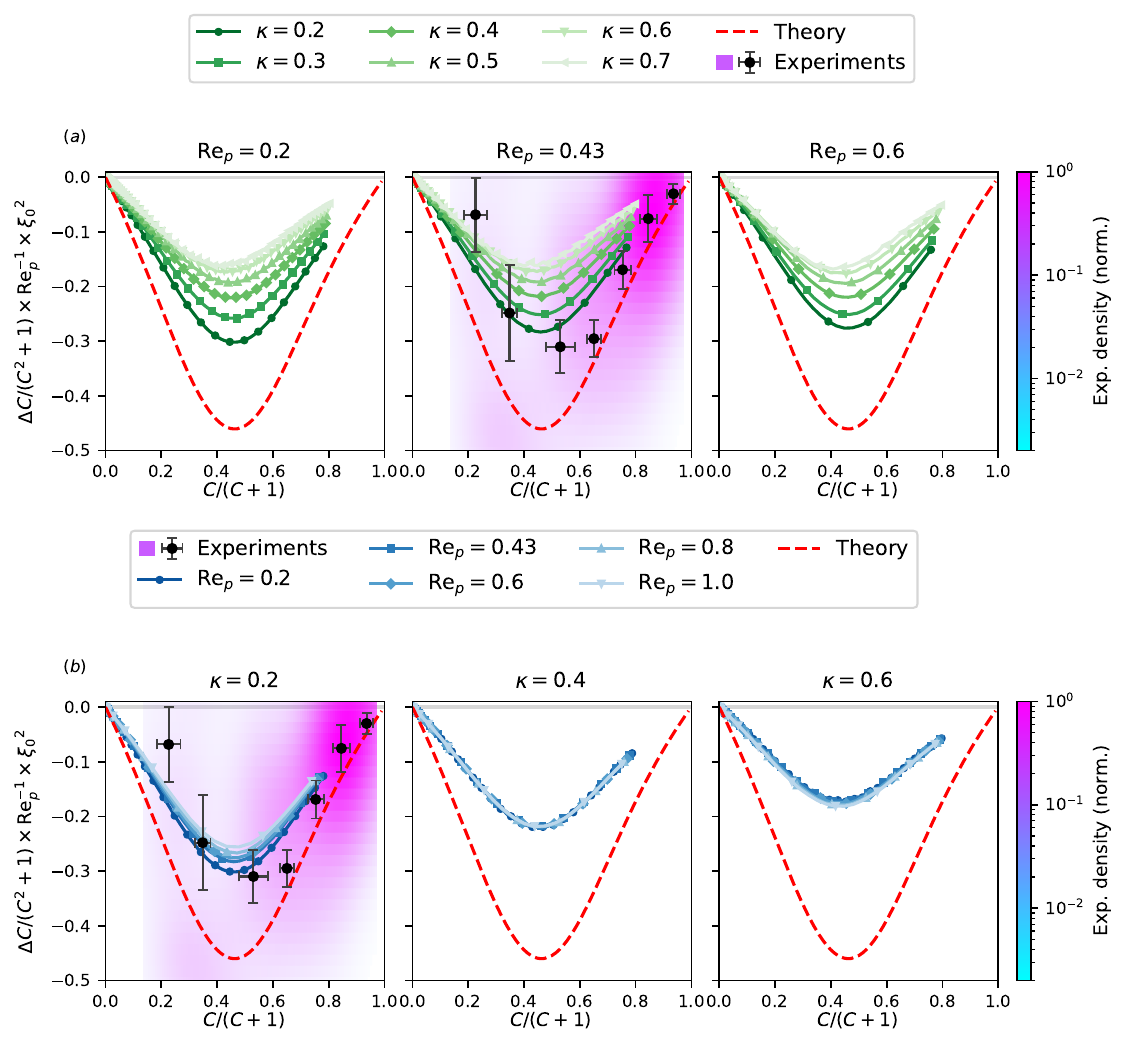}}
    \caption{\label{fig:DCmap}
    Normalized change in the orbit constant in a single Jeffery period, $\Delta C / (C^2+1) \times \text{Re}_p^{-1} \times \xi_0^2$, plotted as a function of $C/(C+1)$ for an oblate spheroid. Note that $C/(C+1)=0$ and $C/(C+1)=1$ correspond to the LR and T modes, respectively. Colored symbols refer to the simulation results at varying $\kappa$, shaded green symbols in Panel (a), or at varying Re$_p$, shaded blue symbols in Panel (b).
    For the experimental condition $\kappa=0.2$ and $\mathrm{Re}_p=0.43$, the shaded background shows a Gaussian kernel-density estimate (KDE) of the single-period drift observations for all the experimental trajectories, evaluated using the same definition of $\Delta C$ as in the simulations and pooled over all experimental runs. The logarithmic colour scale represents the local density of observations normalised by its peak value. The individual-period measurements exhibit scatter, while the systematic drift becomes more clearly resolved through ensemble averaging. Black symbols: mean of these measurements in adaptive bins of $C/(C+1)$ (bins merged until each holds at least 10 periods); vertical bars give the standard error of the bin mean, horizontal bars the standard deviation of $C/(C+1)$ within the bin. This KDE layer visualizes the distribution and spread of the individual-period observations, complementing the binned mean and errorbars by revealing the underlying experimental variability.
    The red dashed line shows the theoretical prediction  for an unbounded weakly inertial spheroid with $\lambda=0.56$ \citep{einarsson2015a,dabade2016effect}.
    See supplementary materials for the directory of the figure including the data and the Jupyter notebook,
    \href{https://cocalc.ai/share/notebook/figure_4/figure_4.ipynb}{https://cocalc.ai/share/notebook/figure\_4/figure\_4.ipynb}.
    }
 \end{figure}

The effects of confinement and inertia can be examined further by avoiding the ambiguity of an arbitrary initial condition.
This analysis can be performed by considering the behaviour of the normalised change in the orbit constant in a single Jeffery period, as done in \citet{dabade2016effect}, see also details in \citet{digiusto2024influence}.
Changes in the orbit constant can be calculated as discrete differences $\Delta C$ over each period of rotation, which we computed using the simulation results.
Positive increments of the orbit constant indicate that the particle is drifting towards a T orbit ($C = \infty$, i.e.
$C/(C + 1) = 1$) whereas negative increments indicate that the particle is attracted to the vorticity-aligned LR orbit ($C= 0$, i.e., $C/(C + 1) = 0$).

Our results for the case of oblate particles with aspect ratio $\lambda =0.56$, are shown in figure\,\ref{fig:DCmap}, where the normalised change in the orbit constant in a single period, $\Delta C/(C^2+1)\text{Re}_p^{-1} \xi_0^2$, is plotted against the normalized orbit constant $C/(C+1)$ for all the simulations.
To keep the drift finite in the near-sphere limit, $\xi_0 \rightarrow \infty$, the normalisation for $\Delta C$ also uses the factor $\xi_0$, defined as $\xi_0 = \sqrt{1/(1-\lambda^2)}$ for an oblate particle.
We compare the orbit drift measured in the simulations with the theoretical prediction. Here we show Eq. (5.19) from \citet{dabade2016effect}. The same curve (within numerical precision) is obtained using Eq.~\eqref{eq:einarsson} and \eqref{Jefferey}.

Figure\,\ref{fig:DCmap} clearly shows that inertia and confinement cause the oblate spheroid to drift towards the LR mode starting from an arbitrary initial orientation, given here by the first right-end point of each colored curve.
This is in agreement with previous theoretical studies, which have shown that, while prolate spheroids are only pushed towards the T orbit, oblate spheroids drift towards a steady LR orbit regardless of their initial orientation provided that their aspect ratio is greater than a critical value of approximately 0.14 \citep{einarsson2015a}.
Only below such critical value, a bifurcation in the orientation dynamics can be observed and distinct basins of orientations that asymptote to the LR and T modes appear, separated by a critical value $C^* \simeq \sqrt{35}$, i.e. $C^*/(C^* + 1) \simeq 0.86$ as predicted by \cite{dabade2016effect}.
The lack of points near or above this critical value in figure\,\ref{fig:DCmap} indicates precisely that the particles never drift toward a full T orbit, even when the initial condition imposed in the simulations is close to that of a T orbit.

The influence of confinement is clearly visible in the plots of panel (a), where the curves associated to different confinement ratios, from $\kappa = 0.2$ (weak confinement) to $\kappa = 0.7$ (strong confinement), are shown at fixed Re$_p$. In all cases, as $\kappa$ increases, the normalized increments become smaller over the range of $C/(C+1)$ covered by the simulations and the curves flatten out, shifting progressively apart from the theoretical predictions. The latter are always found to predict a stronger drift for an unconfined viscous shear flow. Overall, confinement appears to weaken the intensity of the orbit drift and stabilize the orbit executed by the particles.
The plots shown in panel (b) show the influence of inertia at a given confinement ratio.
While the curves are slightly spread at a low confinement ratio ($\kappa = 0.2$), they collapse (as expected) at a high confinement ratio ($\kappa = 0.7$).
This suggests a saturation of the inertial effect above a certain threshold, which can also be appreciated in the inset of figure\,\ref{fig:drift}(b).
A direct comparison with the experimental drift measurements is provided in the second panel of Fig.\,\ref{fig:DCmap}(a) and first panel of Fig.\,\ref{fig:DCmap}(b). For the experimental condition $\kappa=0.2$ and $\mathrm{Re}_p=0.43$, the single-period drift measurements are binned according to $C/(C+1)$ and compared with the corresponding simulation results and the unbounded theoretical prediction. The binned experimental mean (black circles) exhibits a predominantly negative drift over the measured range of $C$, consistent with the direction of the drift predicted by both the simulations and the unbounded theory. The individual experimental measurements display considerable scatter, as illustrated by the KDE background, while the binned mean and its standard error reveal a clearer systematic trend. Despite the larger experimental variability and the comparatively sparse sampling, no systematic positive drift of $C$ is observed. Thus, within the experimentally accessible range, the measurements support a drift toward lower-$C$ orbits rather than an attraction toward the tumbling state for the present aspect ratio, $\lambda=0.56$.
\section{Discussions and concluding remarks}
\label{sec:conclusion}

In this study, we have revisited the angular dynamics of a weakly inertial (Re$_p \leq 1$), neutrally buoyant, moderately oblate spheroid in simple shear flow. The central question was whether the log-rolling (LR) orbit is the only stable long-time orbit for the particle considered here, or whether the tumbling (T) orbit may also appear stable under the experimental conditions. This question was motivated by the apparent discrepancy reported in \citet{digiusto2024influence}, where one trajectory of an oblate spheroid with aspect ratio $\lambda\simeq0.56$ and Re$_p\simeq0.43$ appeared to remain close to the T orbit, although weak-inertia theory predicts attraction to LR for this aspect ratio.

Our results show that, for this particle, LR is the stable long-time orbit and T is unstable. This conclusion is supported by the fully resolved simulations, the theoretical predictions, and the longer experimental time series.
The earlier experimental observations \citep{digiusto2024influence} and the weak-inertia predictions \citep{einarsson2015a,einarsson2015b,dabade2016effect} can be interpreted consistently by accounting for confinement, finite inertia, sensitivity to the initial orientation, and finite observation time. In particular, the experimentally estimated drift is predominantly directed toward lower-$C$ orbits and is quantitatively closer to the confined simulations than to the unbounded theory. Although the finite duration and scatter of the experimental measurements prevent them from establishing the asymptotic orbit independently, they provide no systematic evidence of attraction toward tumbling.

The comparison between theory, simulations, and experiments clarifies why the drift can be difficult to observe. The unbounded weak-inertia theory predicts a faster drift toward LR than the confined simulations. The simulations show that confinement slows the drift toward LR, whereas increasing Re$_p$ tends to accelerate it. Within the parameter range considered here, the influence of confinement on the drift rate is more pronounced than the corresponding increase produced by Reynolds-number effects. As a result, the experimental trajectories, which are necessarily confined, evolve much more slowly than predicted by the unbounded theory. This provides a natural explanation for why trajectories starting close to T may remain apparently tumbling-like during a short experimental time window.

The sensitivity to the initial condition is another important factor. Because the drift rate is smallest near the tumbling orbit, even modest uncertainties in the initial orientation may lead to appreciable differences in the subsequent trajectories. This sensitivity is visible in the comparison of the time series of the orientation components ($n_1$, $n_2$, $n_3$) in figure~\ref{fig:comparison}. It also explains why a direct comparison of orientation time series can be misleading unless the initial orientation and the observation time are carefully accounted for.

The role of confinement is further quantified through the stability exponent of the LR orbit and through the change in the orbit constant over one Jeffery period. The near-collapse of $\gamma_{\text{LR}}/\text{Re}_p$ as a function of $\kappa$ for different Re$_p$ values shows that confinement systematically reduces the drift rate. The same trend is observed in the map of the orbit-constant change, where even moderate confinement produces a clear deviation from the unbounded theory. Increasing confinement weakens the orbit drift, whereas increasing inertia leads to a saturation of the drift intensity.

We have also stressed the importance of the observation time for particles starting near the T orbit in experiments, which is illustrated by the longer experimental runs: for example, experiment 11 (Fig.\,\ref{fig:app_7}), $n_3$ remains almost unchanged up to $t\dot{\gamma}\simeq80$ and increases only afterwards. If the observation had stopped earlier, as in RUN10 of \citet{digiusto2024influence}, the trajectory could have been incorrectly interpreted as tumbling-stable.

These results also underline the importance of boundary conditions for the angular dynamics of non-spherical particles. Wall confinement is distinct from particle-particle hydrodynamic interactions, but both can modify the angular motion by altering the flow disturbance around the particle and by breaking the assumptions of dilute, unbounded theories. Therefore, care is required when applying dilute-limit predictions to confined experiments or to more concentrated suspensions.

The present interpretation is specific to neutrally buoyant particles in simple shear flow. Other effects may alter the angular stability, for example the slip-induced fluid inertial torque produced by sedimentation.
In this case, the angular motion of the particle depends on the direction of gravity relative to the shear plane.
When gravity acts along the flow or vorticity directions, the behaviour remains similar to neutrally buoyant particles.
However, for gravity along the shear direction and sufficiently large sedimentation Reynolds numbers, particles tend to orient upstream, with an angular drift toward the vorticity axis \citep{subramanian2005}.
These subtleties do not affect our interpretation of the experimental results, where the particles were placed in the centre of the simple shear, ensuring that slip is negligible; the conditions are given by \citet{candelier2019time}.
We further note that simple shear flow is a special case amongst all steady linear flows, satisfying $O_{ij}O_{jk}= -S_{ij}S_{jk}$, and $S_{ij}O_{jk} = -O_{ij}S_{jk}$. For other steady linear flows -- such as pure straining (extensional) and solid-body rotational flows -- there are additional terms in equation\,\eqref{eq:einarsson} \citep{candelier2015}, yielding different conclusions concerning the effect of fluid (and particle) inertia.
In a solid-body rotation, only particle inertia affects the angular velocity of the spheroid and T is stable for prolate particles while LR is stable for oblate nearly spherical particles.
In extensional flow, prolate and oblate particles orient their axes of symmetry in the flow plane and eventually reach fixed orientations.
The effects of fluid inertia in other linear flows, however, remain largely unexplored.

The results discussed in this paper represent an advancement of knowledge because they shed light on the stability of the angular dynamics of non-spherical particles at moderate fluid inertia, which is fundamental to understanding the behaviour of suspensions in practical flows.
Many applications -- from industrial processes including paper and pulp production \citep{lundell2011fluid} and combustion soot emission \citep{moffet2009situ} to dispersion systems related to foods, cosmetics, and pharmaceuticals products and particle sorting \citep{erni2009continuous,martel2014inertial} -- operate at Reynolds numbers where fluid-inertia effects are non-negligible.
Clarifying the angular stability in this regime
helps interpret finite-time experiments and shows that confinement must be accounted for when comparing experiments with unbounded weak-inertia theory. A natural next step in future is to examine how this drift-rate modification changes when particles are allowed to translate or interact hydrodynamically with other particles.

\backsection[Acknowledgements]{This work is supported by the Netherlands Organization for Scientific Research (NWO) through the use of supercomputer facilities (Snellius) under Grant No. 2023.026, 2025.008. This publication is part of the project ``Shaping turbulence with smart particles'' with Project No. OCENW.GROOT.2019.031 of the research program Open Competitie ENW XL which is (partly) financed by the Dutch Research Council (NWO). BM thanks F. Candelier for discussions. The work of BM was in part supported by VR grant no. 2021-4452, and BM kindly acknowledges support from the Eindhoven Artificial Intelligence Systems Institute (EAISI) of TU/e for support with a visiting professorship at TU/e.
D.D.G. acknowledges the Universit\'a Italo-Francese, Bando Vinci 2021, cap. 2, progetto C2-257555 ``Fibre flessibili in flusso turbolento ad elevato numero di Reynolds'' for the generous funding during the execution of the experiments discussed in this manuscript.
}

\backsection[Declaration of interests]{The authors report no conflict of interest.}

\backsection[Data availability statement]{The experimental recordings that support the findings of this study are openly available at \href{https://huggingface.co/datasets/ddg93/LeRing_JFM_experiments/tree/main}{LeRing JFM experiments}.
Supplementary movies are available. Computational Notebook can be found online at
\href{https://cocalc.ai/share/notebook}{https://cocalc.ai/share/notebook}.
}

\backsection[Author ORCIDs]{\\
Z. Wang, \href{https://orcid.org/0009-0003-4841-2952}{https://orcid.org/0009-0003-4841-2952};\\
X. M. de Wit, \href{https://orcid.org/0000-0002-7731-0598}{https://orcid.org/0000-0002-7731-0598};\\
D. Di Giusto, \href{https://orcid.org/0000-0003-4413-2454}{https://orcid.org/0000-0003-4413-2454};\\
L. Bergougnoux, \href{https://orcid.org/0000-0002-2988-4394}{https://orcid.org/0000-0002-2988-4394};\\
\'E. Guazzelli, \href{https://orcid.org/0000-0003-3019-462X}{https://orcid.org/0000-0003-3019-462X};\\
C. Marchioli, \href{https://orcid.org/0000-0003-0208-460X}{https://orcid.org/0000-0003-0208-460X};\\
B. Mehlig, \href{https://orcid.org/0000-0002-3672-6538}{https://orcid.org/0000-0002-3672-6538};\\
F. Toschi, \href{https://orcid.org/0000-0001-5935-2332}{https://orcid.org/0000-0001-5935-2332}.
}

\bibliographystyle{jfm}
\bibliography{sis}

\newpage
\appendix

\renewcommand{\thesection}{\Alph{section}} %
\setcounter{figure}{0}
\renewcommand{\thefigure}{A\arabic{figure}}
\setcounter{table}{0}
\renewcommand{\thetable}{A\arabic{table}}
\section{Working-fluid preparation and characterisation in experiments}
\label{sec:working_fluid}
The working fluid was prepared by first adjusting the concentration of
citric acid in water to match the fluid density to that of the particle.
UCON\textsuperscript{TM} Lubricant 75-H-90,000 (Dow Chemical) was then
added to adjust the dynamic viscosity and, consequently, the particle
Reynolds number.

At $25^\circ$C, UCON\textsuperscript{TM} Lubricant 75-H-90,000 has a
density of $1092~\mathrm{kg\,m^{-3}}$ and a dynamic viscosity of
$30~\mathrm{Pa\,s}$. At the same temperature, an aqueous citric-acid
solution at the solubility limit ($60\%$) has a density of
$1294~\mathrm{kg\,m^{-3}}$ and a dynamic viscosity of
$17\times10^{-3}~\mathrm{Pa\,s}$.

For each experimental condition, the density and viscosity of the final
working fluid were measured independently. The density was measured using
an Anton Paar DMA35 densimeter, with an uncertainty of
$1~\mathrm{kg\,m^{-3}}$, while the dynamic viscosity was measured using
an Anton Paar MCR501 rheometer, with an uncertainty of
$0.5\times10^{-3}~\mathrm{Pa\,s}$. The relevant information is summarised
in table\,\ref{tab:working_fluid}.

\begin{table}[h]
\centering
\caption{
Properties of the constituents used to prepare the working fluid and measurement methods used to characterise the final mixture.
}
\label{tab:working_fluid}
\begin{tabular}{@{}p{0.43\linewidth}p{0.51\linewidth}@{}}
\toprule
Quantity & Value / method \\
\midrule
UCON type &
UCON\textsuperscript{TM} Lubricant 75-H-90,000 (Dow Chemical) \\
UCON density at $25^\circ$C &
$1092~\mathrm{kg\,m^{-3}}$ \\
UCON dynamic viscosity at $25^\circ$C &
$30~\mathrm{Pa\,s}$ \\
Citric-acid solution &
$60\%$ aqueous solution (solubility limit) \\
Citric-acid solution density at $25^\circ$C &
$1294~\mathrm{kg\,m^{-3}}$ \\
Citric-acid solution dynamic viscosity at $25^\circ$C &
$17\times10^{-3}~\mathrm{Pa\,s}$ \\
Final-fluid density &
Anton Paar DMA35; uncertainty $1~\mathrm{kg\,m^{-3}}$ \\
Final-fluid dynamic viscosity &
Anton Paar MCR501; uncertainty
$0.5\times10^{-3}~\mathrm{Pa\,s}$ \\
\bottomrule
\end{tabular}
\end{table}

The physical properties of the particle and the background fluid used in the experiments are summarised in Table\,\ref{tab:exp_parameters}.
\begin{table}[!htbp]
    \centering
    \caption{Characteristics of the particle used in the experiments. The columns list the particle aspect ratio $\lambda$, half-length along the symmetry axis $b$, half-length perpendicular to that axis $a$, shear rate $\dot{\gamma}$, and Reynolds number.}
    \label{tab:exp_parameters}
    \begin{tabular}{l l c c c c c c c}
    \toprule
    $\lambda$ & $b$ (mm) & $a$ (mm)  & $\dot{\gamma}~(\text{s}^{-1})$ & $\mu~(\text{Pa}\cdot\text{s})$ & $\rho_f$ (kg m$^{-3}$)& Re$_p$ \\
    \midrule
     $0.56 \pm 0.01$ & $1.27 \pm 0.01$ & $2.28 \pm 0.01$ & $3.15$ & $0.046$ & 1230 & 0.43  \\
    \bottomrule
\end{tabular}
\end{table}

\clearpage
\newpage

\setcounter{figure}{0}
\renewcommand{\thefigure}{B\arabic{figure}}
\setcounter{table}{0}
\renewcommand{\thetable}{B\arabic{table}}
\section{Initial particle orientation and its uncertainty in experiments}
\label{sec:initial_orientation}
The particles were not placed at prescribed initial orientations. For each experimental run, the particle was manually introduced near the centre of the camera fields and as close as possible to the neutral plane of the shear flow, and was then pre-sheared for a short time before recording. Repeating this procedure naturally resulted in different initial orientations from run to run.

The particle motion was recorded simultaneously using two Allied Prosilica
GX1910 cameras viewing the flow-gradient and flow-vorticity planes, with a
spatial resolution of approximately 20 pixels/mm in both views. Spatial
calibration was performed using a checkerboard placed inside the measurement
cell. The two viewing directions were arranged to be perpendicular, while a
checkerboard placed inside the measurement cell was used for spatial
calibration and to check possible optical distortion \citep{di2023dinamica}.

The uncertainty relevant to the initial condition in Fig.\,\ref{fig:comparison} was estimated from the orientation-reconstruction procedure. Uncertainties in the measured AABBs were obtained from repeated perturbations of 1000 experimental frames and propagated through the neural network. The resulting standard deviation of the reconstructed $n_3$ component is $\sigma_{n_3}=0.06$, which is used in Fig.\,\ref{fig:comparison} to evaluate the sensitivity of the theoretical trajectories through $n_3\pm\sigma_{n_3}$.

\clearpage
\newpage

\setcounter{figure}{0}
\renewcommand{\thefigure}{C\arabic{figure}}
\setcounter{table}{0}
\renewcommand{\thetable}{C\arabic{table}}
\section{Effect of the lateral periodic domain size}
\label{sec:f}
\begin{figure}[h]
    \centering
    \includegraphics[width=0.99\linewidth]{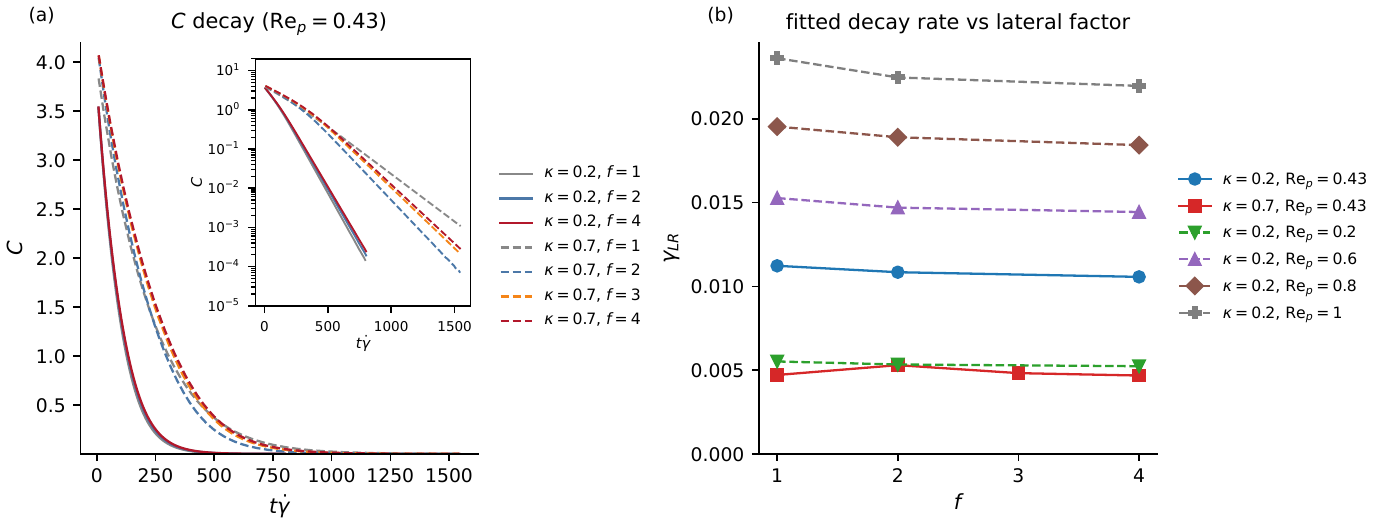}
    \caption{\label{fig:domain_size_test}
    Effect of the periodic domain size on the time evolution of the orbit constant $C$. (a) Time evolution of the orbit constant $C$ for different lateral domain factors $f$ at (Re$_p=0.43$). Solid lines correspond to ($\kappa=0.2$), and dashed lines to ($\kappa=0.7$). The inset shows the same data on a logarithmic vertical scale.
    (b) Fitted log-rolling decay rate ($\gamma_{\rm LR}$) as a function of $f$. Results at ($\kappa=0.2$) are shown for Re$_p=0.2, 0.43, 0.6, 0.8$, and $1.0$, while the ($\kappa=0.7$) series corresponds to (Re$_p=0.43$). Increasing $f$ modifies the quantitative decay rate for smaller $f$, while increasing $f$ even further only poses a minor influence on the decay rate.
    In all the tested cases, $C$ decreases toward zero, indicating drift toward the log-rolling orbit.See supplementary materials for the directory of the figure including the data and the Jupyter notebook,
    \href{https://cocalc.ai/share/notebook/appendix/appendix_fig.ipynb}{https://cocalc.ai/share/notebook/appendix/appendix\_fig.ipynb}.
    }
    \end{figure}

    To assess whether the periodic images in the transverse and flow directions affect the orbit drift, we performed an additional domain-size convergence study in which only the periodic lengths in the flow and spanwise directions were increased, while the wall-normal gap, particle size and aspect ratio, Reynolds number, grid spacing, shear rate, and initial orientation were kept unchanged.
    We define the lateral domain factor $f$ such that the periodic lengths in the flow and spanwise directions are $f$ times those of the cubic domain. Thus, $f=1$ corresponds to the cubic cell, whereas $f=2$, $3$, and $4$ progressively move the periodic images farther away without changing the wall-normal confinement characterised by ($\kappa$).
    Figure\,\ref{fig:domain_size_test}(a) shows the decay of the orbit constant $C$ for different $f$ at experimental Reynolds number Re$_p=0.43$, while Fig.\,\ref{fig:domain_size_test}(b) reports the fitted decay rate $\gamma_{\rm LR}$ for varying Re$_p$ and $\kappa$.
    At $\kappa=0.2$, relative to the $f=4$ result, the cubic domain overestimates $\gamma_{\rm LR}$ by approximately $5$--$8\%$ over Re$_p=0.2$--$1.0$, whereas the $f=2$ domain reduces the difference to approximately $1.8$--$2.6\%$.
    For the experiment-matched condition, ($\kappa=0.2$) and (Re$_p=0.43$), the decay rate at $f=2$ differs from the $f=4$ value by only $2.6\%$. Moreover, the rotation period is essentially unchanged: $T_r/T_J=1.0112$ and $1.0108$ for $f=2$ and $4$, respectively, indicating that the experiment-matched simulations are already only weakly affected by the periodic images at $f=2$. We therefore use $f=2$ for the simulations directly compared with the experiments.
    We also examined the more strongly confined case, $\kappa=0.7$ at Re$_p=0.43$, for which the periodic-image effect is more pronounced. As shown by the dashed lines in Fig.\,\ref{fig:domain_size_test}(a) and the red squares in Fig.\,\ref{fig:domain_size_test}(b), the decay rate exhibits a finite-domain dependence for smaller $f$, with deviations relative to $f=4$ of approximately $+0.7\%$, $+13.3\%$, and $+3.1\%$ for $f=1$, $2$, and $3$, respectively. The apparent agreement between $f=1$ and $f=4$ at this particular confinement should not be interpreted as convergence of the cubic domain. Instead, it illustrates why convergence must be assessed using a systematic sequence of increasing lateral-domain sizes. To minimise the influence of periodic images in the quantitative parameter scans, we therefore use $f=4$ for all simulations reported in the Figs.~3 and 4 of the main text.
    Importantly, enlarging the periodic domain does not change the qualitative orbit selection: in all tested cases, the particle continues to drift toward the log-rolling orbit. This confirms that the qualitative stability selection is not an artifact of the periodic images.

\clearpage
\newpage

\setcounter{figure}{0}
\renewcommand{\thefigure}{D\arabic{figure}}
\setcounter{table}{0}
\renewcommand{\thetable}{D\arabic{table}}
\section{Numerical simulations parameter settings}
\label{sec:sim_param}
The parameter settings used in the numerical simulations are summarised in Table\,\ref{tab:sim_parameters}. The reference case corresponding to the experimental configuration uses $\lambda=0.56$, Re$_p\simeq0.43$, $\kappa=0.2$ and $f=2$.
\begin{table}[!ht]
\centering
\small
\setlength{\tabcolsep}{3pt}
\renewcommand{\arraystretch}{1.08}
  \caption{\label{tab:sim_parameters}
     Parameters of the simulations. In every case $\lambda=b/a=0.56$ and $\rho_p/\rho_f=1$, so that $\mathrm{St}=\mathrm{Re}_p$. Lengths are normalised by the volume-equivalent sphere diameter $D_{eq}=2a\lambda^{1/3}$; $L_1$, $L_2$, and $L_3$ represent the flow, gradient (wall-normal) and vorticity directions, so that $L_2/D_{eq}=\lambda^{-1/3}/\kappa$, while $f=L_1/L_2=L_3/L_2$ measures the lateral extent of the periodic box. The grid is uniform and isotropic with spacing $\Delta x$, and $t_f$ the total simulated time.
  }
\begin{tabular}{@{}cccccccc@{}}
\toprule
Re$_p$
& $\lambda$
& $\kappa$
& $f$
& $N_1\times N_2\times N_3$
& $L_i/D_{eq}$
& $D_{eq}/\Delta x$
& $t_f\dot\gamma$ \\
\midrule

\multicolumn{8}{l}{\emph{experiment-matched trajectories}} \\

0.43
& 0.56
& 0.2
& 2
& $512\times256\times512$
& 12.13, 6.07, 12.13
& 42.2
& 518 \\

\multicolumn{8}{l}{\emph{production scan (decay rates and drift maps)}} \\

\makecell[c]{0.20, 0.43, 0.60,\\0.80, 1.00}
& 0.56
& 0.2
& 4
& $1024\times256\times1024$
& 24.26, 6.07, 24.26
& 42.2
& 460--806 \\

\makecell[c]{0.20, 0.43,\\0.60}
& 0.56
& 0.3
& 4
& $680\times170\times680$
& 16.18, 4.04, 16.18
& 42.0
& 691--1459 \\

\makecell[c]{0.20, 0.43, 0.60,\\0.80, 1.00}
& 0.56
& 0.4
& 4
& $512\times128\times512$
& 12.13, 3.03, 12.13
& 42.2
& 518--1767 \\

\makecell[c]{0.20, 0.43,\\0.60}
& 0.56
& 0.5
& 4
& $400\times100\times400$
& 9.71, 2.43, 9.71
& 41.2
& 863--2088 \\

\makecell[c]{0.20, 0.43, 0.60,\\0.80, 1.00}
& 0.56
& 0.6
& 4
& $344\times86\times344$
& 8.09, 2.02, 8.09
& 42.5
& 691--2489 \\

\makecell[c]{0.20, 0.43,\\0.60}
& 0.56
& 0.7
& 4
& $296\times74\times296$
& 6.93, 1.73, 6.93
& 42.7
& 1151--3051 \\

\multicolumn{8}{l}{\emph{lateral-domain (periodic-image) test}} \\

0.43
& 0.56
& 0.2
& 1
& $256\times256\times256$
& 6.07, 6.07, 6.07
& 42.2
& 806 \\

\makecell[c]{0.20, 0.43, 0.60,\\0.80, 1.00}
& 0.56
& 0.2
& 2
& $512\times256\times512$
& 12.13, 6.07, 12.13
& 42.2
& 460--2133 \\

0.43
& 0.56
& 0.7
& 1
& $74\times74\times74$
& 1.73, 1.73, 1.73
& 42.7
& 1554 \\

0.43
& 0.56
& 0.7
& 2
& $148\times74\times148$
& 3.47, 1.73, 3.47
& 42.7
& 1554 \\

0.43
& 0.56
& 0.7
& 3
& $222\times74\times222$
& 5.20, 1.73, 5.20
& 42.7
& 1554 \\

\makecell[c]{0.20, 0.60,\\0.80, 1.00}
& 0.56
& 0.2
& 1
& $256\times256\times256$
& 6.07, 6.07, 6.07
& 42.2
& 869--1239 \\

\bottomrule
\end{tabular}

\end{table}

\clearpage
\newpage

\setcounter{figure}{0}
\renewcommand{\thefigure}{E\arabic{figure}}
\setcounter{table}{0}
\renewcommand{\thetable}{E\arabic{table}}
\section{Grid-refinement check}
\label{sec:grid_refinement_check}
\begin{figure}[h]
\centering
\includegraphics[width=0.99\linewidth]{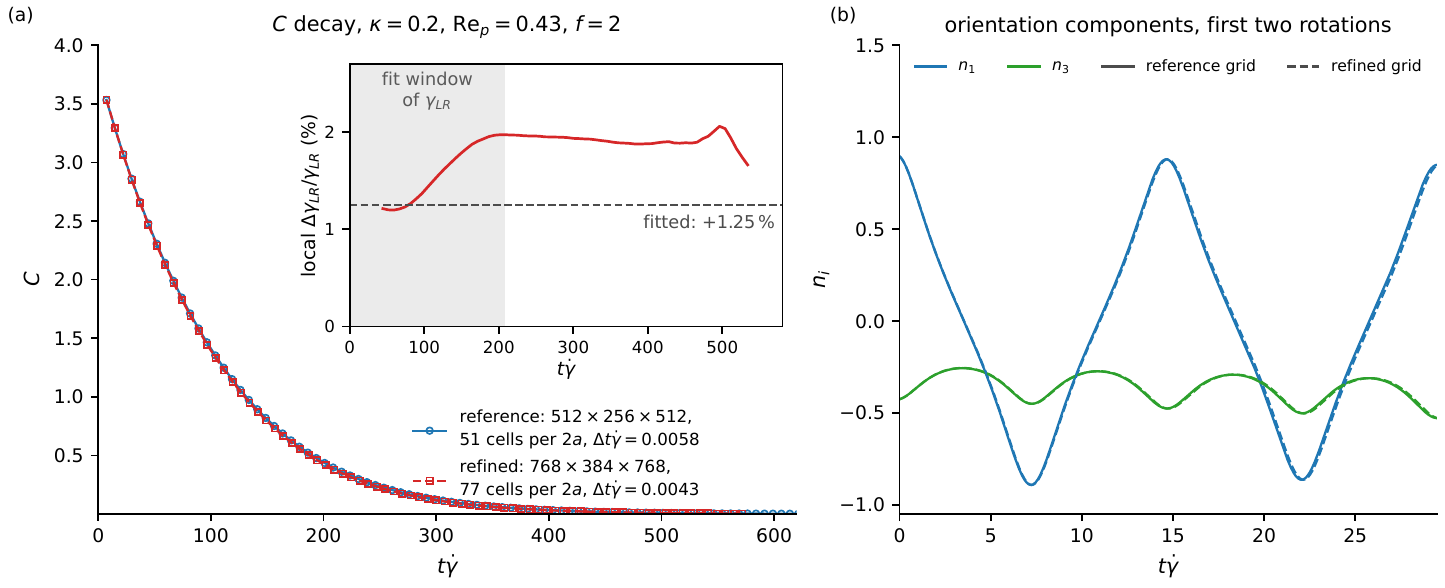}
\caption{\label{fig:SI_resolution}
Grid-refinement check for the experiment-matched configuration
($\kappa=0.2$, Re$_p=0.43$, $f=2$): reference grid
$512\times256\times512$ ($51$ cells across $2a$,
$\Delta t\dot\gamma=0.0058$) versus refined grid
$768\times384\times768$ ($77$ cells across $2a$,
$\Delta t\dot\gamma=0.0043$).
(a) Orbit constant $C(t)$ (one median value per half rotation) for the two
grids; inset: local decay-rate difference
$-\mathrm{d}\ln(C_{\rm refined}/C_{\rm ref})/\mathrm{d}t$ normalised by
$\gamma_{\rm LR}$, with the dashed line showing the difference between the two
fitted rates ($+1.25\%$) and the shaded area indicating the fitting window.
(b) Orientation components $n_1$ and $n_3$ over the first two rotations
(solid: reference; dashed: refined).
See supplementary materials for the directory of the figure including the data and the Jupyter notebook,
\href{https://cocalc.ai/share/notebook/appendix/appendix_fig.ipynb}{https://cocalc.ai/share/notebook/appendix/appendix\_fig.ipynb}.
}
\end{figure}

The production runs use a uniform isotropic grid with
$D_{eq}/\Delta x \simeq 42.2$, corresponding to about $51$ grid cells across
the particle major diameter $2a$, and $\Delta t\dot\gamma \le 0.006$.
To verify that this resolution is adequate, we repeated the experiment-matched
case ($\kappa=0.2$, Re$_p=0.43$, $f=2$) with both the grid spacing and the
time step reduced by a factor $1.5$, using a
$768\times384\times768$ grid ($77$ cells across $2a$), while keeping all other
parameters and the initial orientation unchanged.
Figure~\ref{fig:SI_resolution} compares the reference and refined runs.
The fitted decay rate changes from
$\gamma_{\rm LR}/$Re$_p=0.025217$ to $0.025532$, i.e. by $+1.25\%$.
The rotation period changes by only $0.4\%$, and the orientation traces over
the first two rotations are nearly indistinguishable.
We therefore estimate the discretisation uncertainty of the reported decay
rates to be about $1$--$2\%$, which is much smaller than the confinement effect
and the lateral-domain effect discussed in the main text.

\clearpage
\newpage

\setcounter{figure}{0}
\renewcommand{\thefigure}{F\arabic{figure}}
\setcounter{table}{0}
\renewcommand{\thetable}{F\arabic{table}}
\section{Inertial migration effect}
\label{sec:inertial_mig}
A possible additional effect is the inertial migration of the particle centre across the gap. For neutrally buoyant spheres in wall-bounded simple shear or plane Couette flow, previous theory and simulations show that the mid-plane is a stable equilibrium position at low Reynolds numbers \citep{ho1974inertial,feng1994direct,anand2023inertial}. For non-spherical particles, lateral migration can be coupled to the particle orientation and is less well established. \cite{huang2015motion} studied a neutrally buoyant elliptical particle in a two-dimensional shear flow and found that inertia can induce a transition from continuous tumbling to a fixed stationary orientation only above a critical Reynolds number, estimated as $Re_{cr}\simeq20$--$120$ for aspect ratios between 0.2 and 0.8 and confinement ratios between 0.1 and 0.4. The Reynolds numbers considered here, Re$_p\lesssim1$, are far below this range.
These results support the physical relevance of the mid-plane configuration for a spherical particle under comparable conditions. For a three-dimensional oblate spheroid, however, lateral migration may be coupled to the instantaneous particle orientation, and its magnitude and equilibrium positions are less well established. Fixing the particle centre at the mid-plane should therefore be regarded as a controlled modelling choice that isolates the angular dynamics, rather than as evidence that translational migration is negligible.
Assessing the coupled translational and rotational motion of three-dimensional oblate spheroids remains outside the scope of the present study.

\clearpage
\newpage

\setcounter{figure}{0}
\renewcommand{\thefigure}{G\arabic{figure}}
\setcounter{table}{0}
\renewcommand{\thetable}{G\arabic{table}}
\section{Additional data}
\label{sec:app_data}

\begin{figure}[h]
\centerline{\includegraphics[width=0.99\linewidth]{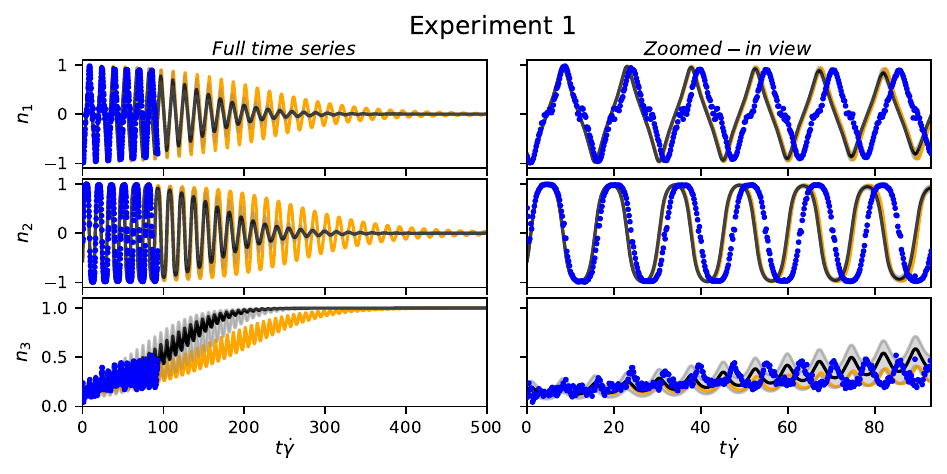}}
    \caption{Comparison of experimental time series (filled symbols) with results of numerical simulations (thick solid lines) and theoretical predictions (thin solid lines) for the case {\tt Experiment 1}. Panels on the left illustrate the full time series, while panels on the right show the corresponding zoomed-in views.
   See the supplementary materials for the figure directory containing the data and Jupyter notebook,
\href{https://cocalc.ai/share/notebook/figure_2/Figure_2.ipynb}{https://cocalc.ai/share/notebook/figure\_2/Figure\_2.ipynb}.
The panels correspond to Experiment 1, as labelled in the Jupyter notebook; the corresponding dynamics are shown in Movie 4.
    }
    \label{fig:app_1}
 \end{figure}

\begin{figure}[h]
    \centerline{\includegraphics[width=0.99\linewidth]{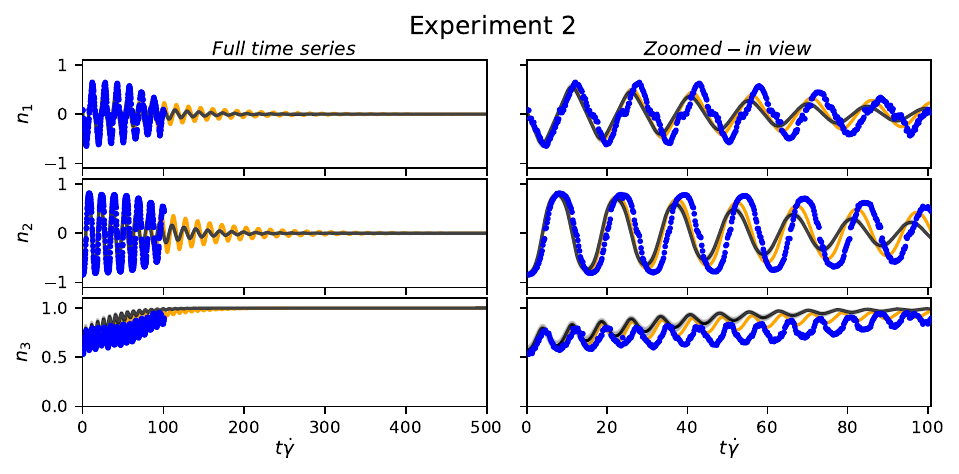}}
    \caption{\label{fig:Appendix_FIGURE_RUN2} Comparison of experimental time series (filled symbols) with results of numerical simulations (thick solid lines) and theoretical predictions (thin solid lines) for the case {\tt Experiment 2}. Panels on the left illustrate the full time series, while panels on the right show the corresponding zoomed-in views.
    See the supplementary materials for the figure directory containing the data and Jupyter notebook,
\href{https://cocalc.ai/share/notebook/figure_2/Figure_2.ipynb}{https://cocalc.ai/share/notebook/figure\_2/Figure\_2.ipynb}.
The panels correspond to Experiment 2, as labelled in the Jupyter notebook; the corresponding dynamics are shown in Movie 5.
    }
\end{figure}

\begin{figure}
    \centerline{\includegraphics[width=0.99\linewidth]{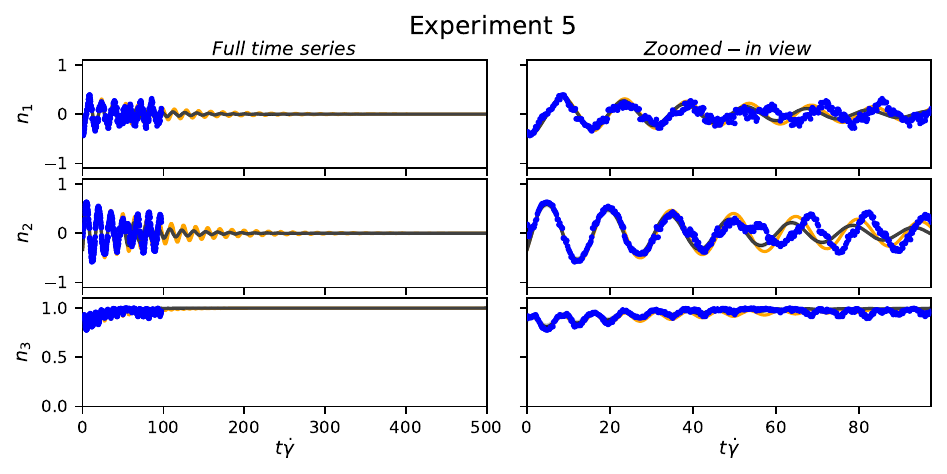}}
    \caption{\label{fig:Appendix_FIGURE_RUN5}
    Comparison of experimental time series (filled symbols) with results of numerical simulations (thick solid lines) and theoretical predictions (thin solid lines) for the case {\tt Experiment 5}. Panels on the left illustrate the full time series, while panels on the right show the corresponding zoomed-in views.
    See the supplementary materials for the figure directory containing the data and Jupyter notebook,
\href{https://cocalc.ai/share/notebook/figure_2/Figure_2.ipynb}{https://cocalc.ai/share/notebook/figure\_2/Figure\_2.ipynb}.
The panels correspond to Experiment 5, as labelled in the Jupyter notebook; the corresponding dynamics are shown in Movie 6.
    }
 \end{figure}

\begin{figure}
    \centerline{\includegraphics[width=0.99\linewidth]{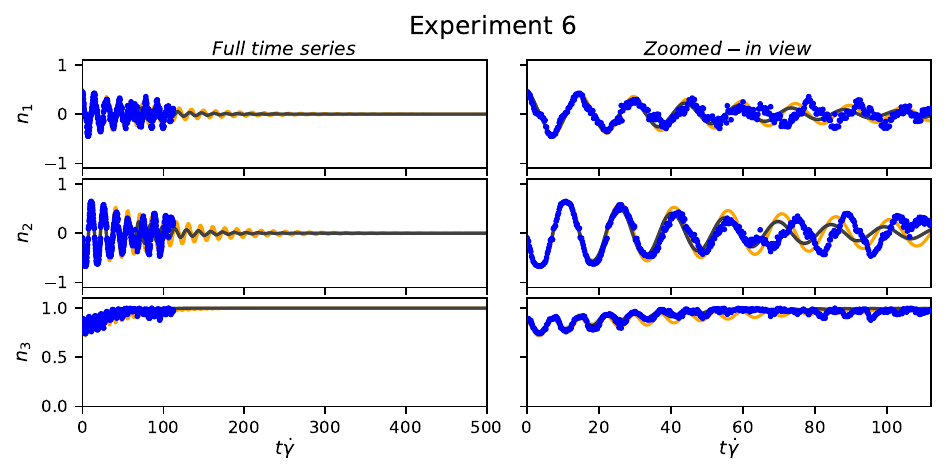}}
    \caption{\label{fig:Appendix_FIGURE_RUN6} Comparison of experimental time series (filled symbols) with results of numerical simulations (thick solid lines) and theoretical predictions (thin solid lines) for the case {\tt Experiment 6}. Panels on the left illustrate the full time series, while panels on the right show the corresponding zoomed-in views.
    See the supplementary materials for the figure directory containing the data and Jupyter notebook,
\href{https://cocalc.ai/share/notebook/figure_2/Figure_2.ipynb}{https://cocalc.ai/share/notebook/figure\_2/Figure\_2.ipynb}.
The panels correspond to Experiment 6, as labelled in the Jupyter notebook; the corresponding dynamics are shown in Movie 7.
    }
 \end{figure}

\begin{figure}
    \centerline{\includegraphics[width=0.99\linewidth]{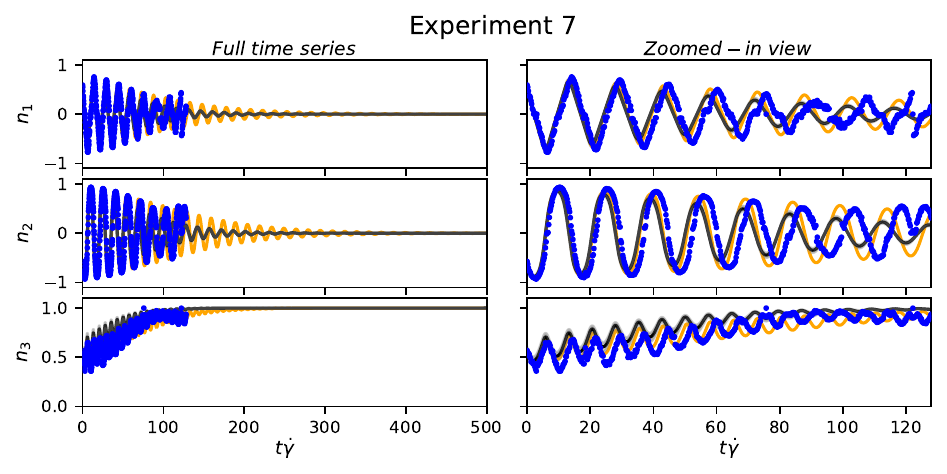}}
    \caption{\label{fig:Appendix_FIGURE_RUN7}
    Comparison of experimental time series (filled symbols) with results of numerical simulations (thick solid lines) and theoretical predictions (thin solid lines) for the case {\tt Experiment 7}. Panels on the left illustrate the full time series, while panels on the right show the corresponding zoomed-in views.
   See the supplementary materials for the figure directory containing the data and Jupyter notebook,
\href{https://cocalc.ai/share/notebook/figure_2/Figure_2.ipynb}{https://cocalc.ai/share/notebook/figure\_2/Figure\_2.ipynb}.
The panels correspond to Experiment 7, as labelled in the Jupyter notebook; the corresponding dynamics are shown in Movie 8.
    }
 \end{figure}

\begin{figure}
    \centerline{\includegraphics[width=0.99\linewidth]{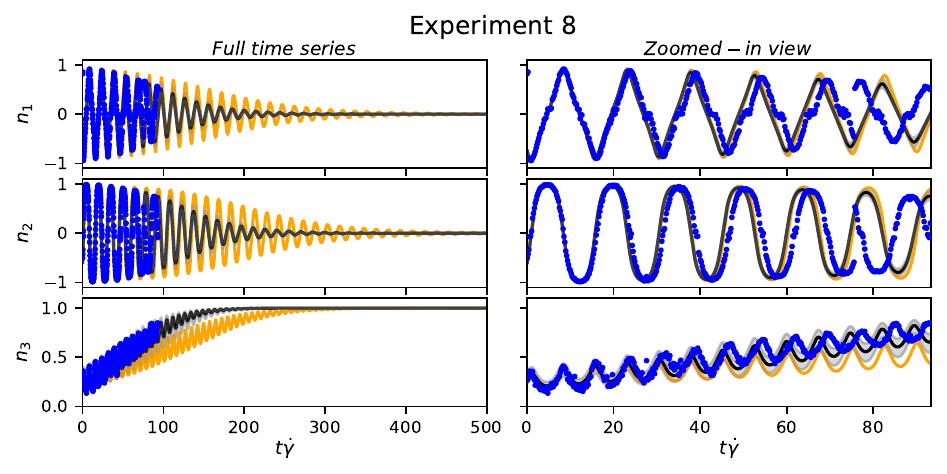}}
    \caption{\label{fig:Appendix_FIGURE_RUN8} Comparison of experimental time series (filled symbols) with results of numerical simulations (thick solid lines) and theoretical predictions (thin solid lines) for the case {\tt Experiment 8}. Panels on the left illustrate the full time series, while panels on the right show the corresponding zoomed-in views.
   See the supplementary materials for the figure directory containing the data and Jupyter notebook,
\href{https://cocalc.ai/share/notebook/figure_2/Figure_2.ipynb}{https://cocalc.ai/share/notebook/figure\_2/Figure\_2.ipynb}.
The panels correspond to Experiment 8, as labelled in the Jupyter notebook; the corresponding dynamics are shown in Movie 9.
    }
 \end{figure}

\begin{figure}
    \centerline{\includegraphics[width=0.99\linewidth]{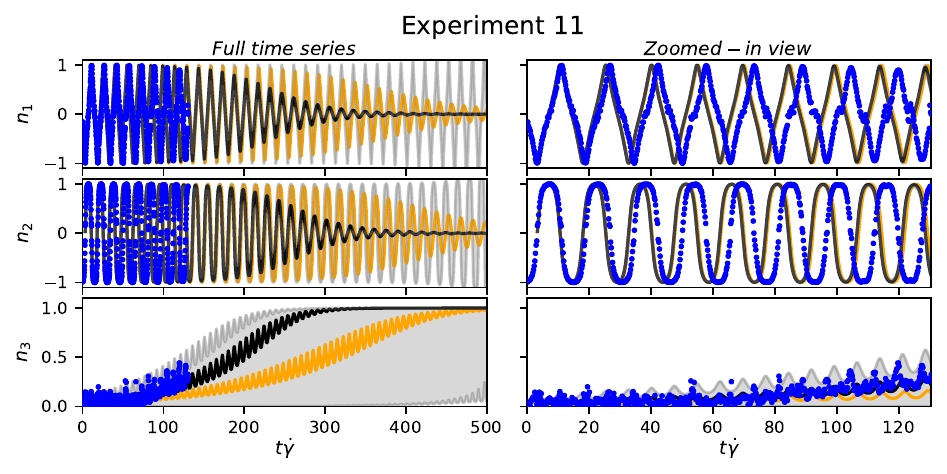}}
    \caption{Comparison of experimental time series (filled symbols) with results of numerical simulations (thick solid lines) and theoretical predictions (thin solid lines) for the case {\tt Experiment 11}. Panels on the left illustrate the full time series, while panels on the right show the corresponding zoomed-in views.
   See the supplementary materials for the figure directory containing the data and Jupyter notebook,
\href{https://cocalc.ai/share/notebook/figure_2/Figure_2.ipynb}{https://cocalc.ai/share/notebook/figure\_2/Figure\_2.ipynb}.
The panels correspond to Experiment 11, as labelled in the Jupyter notebook; the corresponding dynamics are shown in Movie 10.
    }
 \label{fig:app_7}
 \end{figure}

In this Appendix, we present additional comparisons among experiments, theory, and simulations, as shown in figure\,\ref{fig:app_1}-\ref{fig:app_7}. The experiments shown here were carried out using the same particle geometry and shear flow settings as those discussed in figure\,\ref{fig:comparison} of the main text. The parameter settings in the simulations are the same as those in figure\,\ref{fig:comparison} of the main text: Re$_p$ = 0.43, $\lambda=0.56$, $\kappa=0.2$, lateral domain factor $f=2$ and grid $512\times256\times512$ (table\,\ref{tab:sim_parameters}).
The only difference among these runs is the initial orientation of the particle.

\clearpage
\newpage

\setcounter{figure}{0}
\renewcommand{\thefigure}{H\arabic{figure}}
\setcounter{table}{0}
\renewcommand{\thetable}{H\arabic{table}}
\section{Comparison of rotation periods}
\label{sec:Tj_Tr}
\begin{figure}[h]
    \centering
    \includegraphics[width=0.9\linewidth]{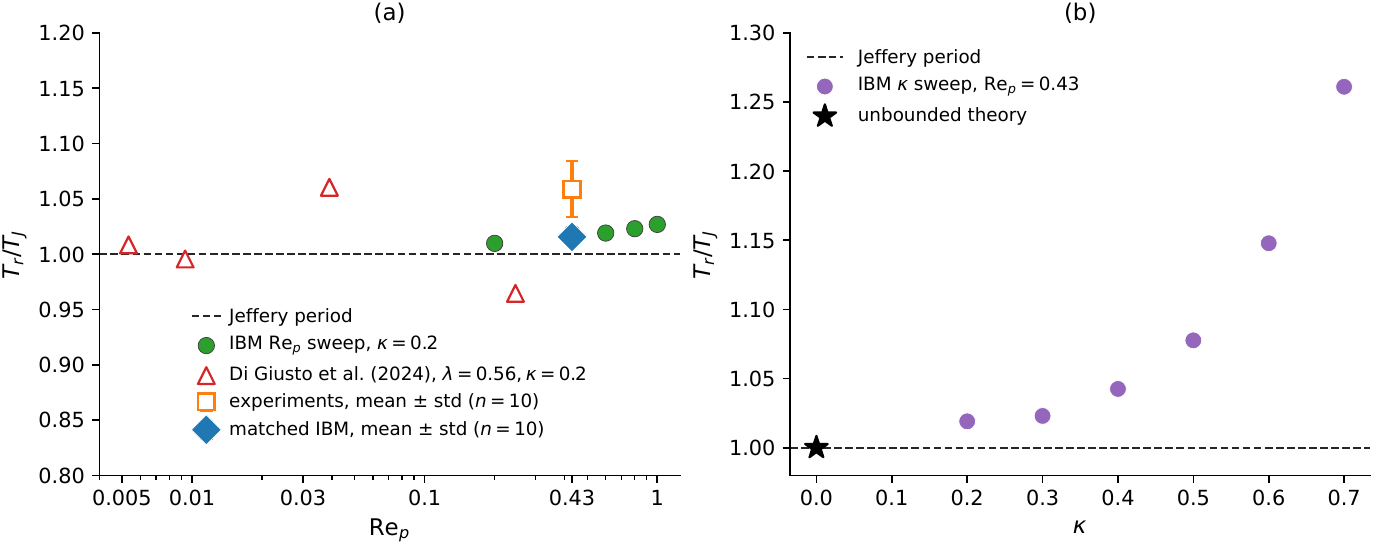}
    \caption{
    Rotation period normalised by the Jeffery period $T_J$.
    (a) Dependence on Re$_p$. Green symbols denote the IBM simulation Re$_p$ sweep at fixed $\kappa=0.2$; red open triangles denote the experimental data from \cite{digiusto2024influence}; the orange square denotes the mean and standard deviation of the present experiments; the blue diamond denotes the mean and standard deviation of the matched IBM simulations; the dashed line denotes the Jeffery period.
    (b) Dependence on $\kappa$ from the IBM $\kappa$ sweep at fixed Re$_p=0.43$. The black star denotes the unbounded-theory value, corresponding to $T_r/T_J=1$.
    }
    \label{fig:SI_period_comparison}
    \end{figure}

In addition to causing drift across Jeffery orbits, inertia and confinement may also affect the rotation period. We therefore compare the period obtained from experiments, simulations, and theory in Fig.\,\ref{fig:SI_period_comparison}. The period is normalised by the Jeffery period $T_J$.
Figure\,\ref{fig:SI_period_comparison}(a) shows the dependence on Re$_p$. The experimental data of \citet{digiusto2024influence} remain scattered around $T_r/T_J=1$ and show no systematic dependence on Re$_p$ for Re$_p\leq1$. The present experimental mean is slightly larger than the matched IBM result and the unbounded theory. The IBM Re$_p$ sweep at fixed $\kappa=0.2$ also shows only a weak dependence on inertia.
\\
Figure\,\ref{fig:SI_period_comparison}(b) shows the dependence on confinement. At fixed Re$_p=0.43$, the normalised period increases from $T_r/T_J\simeq1.02$ at $\kappa=0.2$ to approximately $1.26$ at $\kappa=0.7$. By contrast, at fixed $\kappa=0.2$, it changes only from approximately $1.01$ to $1.03$ over Re$_p=0.2$--$1.0$. Confinement therefore has a much stronger influence on the rotation period than inertia over the parameter range considered here. Hence, in addition to affecting the drift rate across Jeffery orbits, confinement also has a modest effect on the rotation period.
\\
We also examined whether finite inertia and confinement produce a measurable distortion of the particle trajectory relative to a Jeffery orbit. For one representative rotation at an intermediate value of $C$, we determined the Jeffery orbit that best fits each orientation trajectory and calculated the root-mean-square angular deviation from this fitted orbit.
The best-fitting Jeffery orbit was obtained by minimizing the mean squared angular separation between the measured orientation vectors and the Jeffery solution over one complete rotation, allowing the orbit constant and phase to vary.
The resulting deviations are approximately $0.7^\circ$ for the matched IBM simulation, $0.9^\circ$ for the weak-inertia theoretical trajectory, and $2.4^\circ$ for the experiment. The larger experimental value is consistent with the uncertainty of the image-based orientation reconstruction. These small deviations indicate that the trajectories remain approximately Jeffery-like over an individual rotation, while the orbit constant evolves slowly over successive rotations. Within the experimental uncertainty and the numerical parameter range considered here, we therefore find no evidence of a systematic orbit-shape distortion that would affect the conclusions concerning LR and T stability.
    \\

\clearpage
\newpage

\setcounter{figure}{0}
\renewcommand{\thefigure}{I\arabic{figure}}
\setcounter{table}{0}
\renewcommand{\thetable}{I\arabic{table}}
\section{Calculation of the orbit constant $C$}
\label{sec:Ccalc}

The orbit constant associated to the time series, shown in figure\,\ref{fig:comparison}, can be obtained by measuring the periods of the oscillations of $n_3$ over time.
The detailed procedure for calculating $C$ from the particle orientation trajectory $\vec{n}$ is as follows.

\begin{figure}[h]
    \centerline{
    \includegraphics[width=1\linewidth]{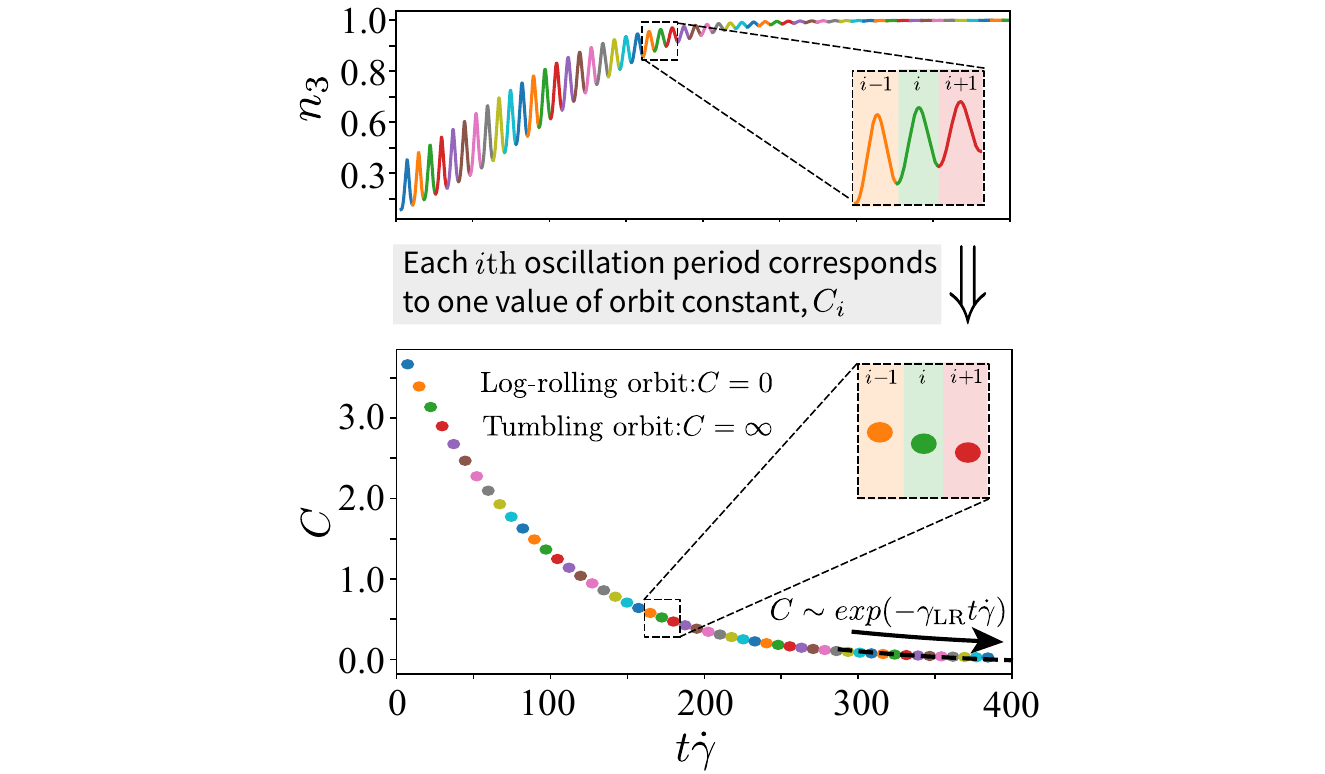}
    }
    \caption{\label{fig:drift_a} Calculation of the orbit constant $C$ (shown here for the case {\tt{Experiment 4}}, chosen as reference): Each value corresponds to one oscillation period of the orientation vector $n_3$. When the particle trajectory drifts toward its LR orbit, $C$ decays exponentially in time as a function of the stability exponent $\gamma_{\text{LR}}$ ($\gamma_{\text{LR}}>0$).
    }
\end{figure}

Starting from the definitions used in the manuscript,
\begin{equation}
n_1=\sin\theta\sin\phi,\qquad
n_2=\sin\theta\cos\phi,\qquad
n_3=\cos\theta ,
\label{eq:response_orientation_components}
\end{equation}
and from the Jeffery-orbit relation
\begin{equation}
\tan\theta =\frac{C\lambda}{\left(\lambda^2\cos^2\phi+\sin^2\phi\right)^{1/2}},
\label{eq:response_jeffery_C}
\end{equation}
we obtain from Eq.~\eqref{eq:response_jeffery_C}
\begin{equation}
C=\frac{\tan\theta}{\lambda}\left(\lambda^2\cos^2\phi+\sin^2\phi\right)^{1/2}.
\label{eq:response_C_theta_phi}
\end{equation}
Using Eq.~\eqref{eq:response_orientation_components}, we have
\begin{equation}
\sin\phi=\frac{n_1}{\sin\theta},\qquad
\cos\phi=\frac{n_2}{\sin\theta}.
\label{eq:response_sin_cos_phi}
\end{equation}
Substituting Eq.~\eqref{eq:response_sin_cos_phi} into the square-root factor in Eq.~\eqref{eq:response_C_theta_phi} gives
\begin{equation}
\left(\lambda^2\cos^2\phi+\sin^2\phi\right)^{1/2}=\frac{\sqrt{n_1^2+\lambda^2 n_2^2}}{\sin\theta}.
\label{eq:response_square_root_factor}
\end{equation}
Finally, substituting Eq.~\eqref{eq:response_square_root_factor} into Eq.~\eqref{eq:response_C_theta_phi}, and using $\tan\theta=\sin\theta/\cos\theta$ together with $n_3=\cos\theta$ from Eq.~\eqref{eq:response_orientation_components}, yields
\begin{equation}
    C=\frac{\sqrt{n_1^2+\lambda^2 n_2^2}}{\lambda |n_3|}.
\label{eq:response_C_from_n}
\end{equation}
The absolute value is used because $C$ is a non-negative orbit constant, and the particle orientation axis is invariant under $\boldsymbol{n}\to-\boldsymbol{n}$.
It can be observed from Fig.\,\ref{fig:drift_a} that $C$ tends to zero with an exponential decay $C \sim exp(-\gamma_{\text{LR}} t \dot{\gamma})$ with $\gamma_{\mathrm{LR}}$ the stability exponent of the LR-orbit \citep{einarsson2015a}. This exponent can be extracted by fitting the exponential decay rate of the orbit constant deviations from the stable LR-orbit (i.e., $C=0$).

The experimental trajectories are reanalysed in terms of the orbit constant $C$ (Fig.\,\ref{fig:drift}) and the period-to-period drift $\Delta C$ (Fig.\,\ref{fig:DCmap}).
The representation of $\Delta C$ allows the drift to be compared as a function of the instantaneous orbit constant, rather than for a single initial condition.
For consistency with the numerical analysis, the orbit constant is first represented by its median value over each half rotation, and the period-to-period drift $\Delta C$ is evaluated using the centred difference
\begin{equation}
    \Delta C_i=C_{i+1}-C_{i-1}.
\end{equation}
This procedure is applied to all available experimental trajectories, which correspond to different initial orientations.

\clearpage
\newpage
\setcounter{figure}{0}
\renewcommand{\thefigure}{C\arabic{figure}}

\end{document}